\documentclass{article}

\usepackage[letterpaper,top=2cm,bottom=2.5cm,left=3cm,right=3cm,marginparwidth=1.75cm]{geometry}

\usepackage{amsmath}
\usepackage{amsfonts}
\usepackage{amsthm}
\usepackage{amssymb}
\usepackage{algorithm}
\usepackage{algpseudocode}
\usepackage{authblk} 
\usepackage{enumitem}
\usepackage{multirow}
\usepackage{colortbl}
\usepackage{color}
\usepackage[table]{xcolor}
\usepackage{epigraph}
\usepackage{graphicx}
\usepackage{caption}
\usepackage{subcaption}
\usepackage{hyperref}
\usepackage{tabularx}
\usepackage{float}
\usepackage{longtable}
\usepackage{pdfpages}
\usepackage{tabularx}
\usepackage{pdflscape}
\usepackage[acronym,toc]{glossaries}
\usepackage{setspace}
\usepackage{mathtools}
\usepackage{tikz}
\usepackage{url}
\usetikzlibrary{calc}

\usepackage{mathtools}

\newcommand{\ket}[1]{\displaystyle | #1 \rangle}
\newcommand{\bra}[1]{\displaystyle \langle #1|}

\newcommand{\di}[1]{\displaystyle \,\mathrm{d} #1}

\newtheorem{theorem}{Theorem}[section]
\theoremstyle{definition}
\newtheorem{definition}{Definition}[section]

\title{Stochastic quantum adiabatic algorithm with fractional Brownian motion}

\author[1]{J.A.D.O. Chinthila\thanks{osanda.chinthila@gmail.com}}
\author[2]{B.P.W. Fernando\thanks{Corresponding author: bpw@sci.sjp.ac.lk}}
\author[1]{A.C. Mahasinghe\thanks{anuradhamahasinghe@maths.cmb.ac.lk}}
\author[2]{R.P.K. De Silva\thanks{kaushika@sjp.ac.lk}}
\author[2]{K.K.W.A.S. Kumara\thanks{sarath@sjp.ac.lk}}

\affil[1]{Department of Mathematics, University of Colombo, Sri Lanka}
\affil[2]{Department of Mathematics, University of Sri Jayewardenepura, Sri Lanka}
\date{April 28, 2025}

\begin{document}

\maketitle

\begin{abstract}
    Adiabatic Quantum Computing relies on the quantum adiabatic theorem, which states that a quantum system evolves along its ground state with time if the governing Hamiltonian varies infinitely slowly. However, practical limitations force computations to be performed within limited times, exposing the system to transitions into excited states, and thereby reducing the success probability. Here we investigate the counterintuitive hypothesis that incorporating stochastic noise, specifically noise driven by \textit{fractional Brownian motion}, in a \textit{non-Markovian} setup can enhance the performance of adiabatic quantum computing by improving its success probability at limited evolution times. The study begins by developing the mathematical framework to introduce stochastic noise multiplicatively into the Schr\"{o}dinger equation, resulting in a stochastic Schr\"{o}dinger equation. To preserve It\^{o} integrability within the non-Markovian framework, a semimartingale approximation for fractional Brownian motion is employed. We perform numerical simulations to compare the performance of the quantum adiabatic algorithm with and without noise driven by fractional Brownian motion using the NP-complete Exact Cover-3 problem, transformed into the Ising model. Our results exhibit an improvement in success probability in the presence of noise driven by fractional Brownian motion with Hurst parameter $0<H<\frac{1}{2}$ and an increase in speedup as $H$ approaches 0. Although simulations are limited to problems involving a modest number of qubits, evidence suggests that the proposed approach scales favorably with the system size.
\end{abstract}


\section{Introduction}
\label{sec: Introduction}

Adiabatic quantum computing was first proposed in 2000 by Farhi et al. \cite{farhi2000quantum} and has gained significant attention due to its potential advantages in areas such as combinatorial optimization \cite{farhi2001quantum,farhi2000numerical}, quantum annealing \cite{finnila1994quantumannealing,santoro2006quantumannealing}, quantum search \cite{roland2002quantumsearch} and period finding \cite{hen2014periodfinding}. This model has been proven to be equivalent to the circuit model of quantum computing with polynomial overhead \cite{aharonov2008adiabatic}. The foundation for adiabatic quantum computing is laid out by the adiabatic theorem of quantum mechanics, which states that if a quantum system is initialized in its ground state, as long as its Hamiltonian is varied slowly enough, the system is likely to remain in the instantaneous ground state throughout the evolution. This allows adiabatic quantum computing to solve optimization problems by encoding the solution in the ground state of the final Hamiltonian.

However, the adiabatic theorem is inherently asymptotic, meaning it guarantees that the system will remain in the ground state with probability 1, only in the limit of infinitely slow evolution. In practical applications, infinitely long evolution times are infeasible, as quantum systems are highly sensitive to decoherence and other interactions with the environment. Consequently, computations must be completed within a finite evolution time, leading to non-negligible probabilities for transitions into excited states \cite{grant2020adiabatic}. This deviation from ideal adiabatic evolution reduces the likelihood of obtaining the correct solution at the end of the computation and makes it essential to explore methods of improving the success probabilities of adiabatic quantum computing.

\par The standard quantum adiabatic algorithm can be stated as follows. The algorithm operates by initializing the system in the easily prepared ground state of a simple initial Hamiltonian $H_I$. The final Hamiltonian $H_F$ is defined such that its ground state encodes the solution to the problem. Next, the adiabatic path which gradually transforms the initial Hamiltonian to the final Hamiltonian is declared by defining the time dependent Hamiltonian $H(t) = (1-s(t))H_I + s(t)H_F$, where $s(t)$ varies from $0$ to $1$ over the evolution time $T$. The time evolution of the system takes place according to the Schr\"{o}dinger equation
\begin{equation}
    \di{\ket{\psi(t)}} = -\imath H(t)\ket{\psi(t)}\di{t}
\end{equation}
with the slowly varying time dependent Hamiltonian $H(t)$. At the end of time evolution, the state of the system $\ket{\psi(T)}$ is measured, which will be close to the ground state of the final Hamiltonian.

Several studies have proposed novel strategies to improve the standard quantum adiabatic algorithm. Cao et al. demonstrated that introducing an additional catalytic Hamiltonian can significantly improve success probability of the algorithm by modifying the energy landscape of the system \cite{cao2021speedup}. Crosson et al. presented numerical evidence suggesting that initializing the system in excited states and adding a random local Hamiltonian to the middle of evolution can enhance performance for the hardest problems in certain combinatorial optimization problem ensembles \cite{crosson2014different}. An early research on the robustness of adiabatic quantum computing suggests that unitary perturbations, often considered sources of error, may actually offer opportunities to speed up the quantum adiabatic algorithm \cite{childs2001robustness}. That being said, this area has been rarely investigated. One notable exception particularly relevant to our work is the investigation of ground state preparation in the Bose-Hubbard model in the presence of stochastic noise by Xu \cite{xu2018adiabatic}.  This study suggests that the noise driven by Brownian motion (white noise), added proportionally to time dependent system Hamiltonian, could counterintuitively enhance the probability of success in the preparation of the ground state.

\par Our research extends this counterintuitive idea by investigating the effects of non-Markovian noise driven by fractional Brownian motion on the performance of the quantum adiabatic algorithm. Unlike previous studies that employed Markovian processes, fractional Brownian motion represents a family of Gaussian processes characterized by a memory parameter known as the Hurst exponent ($H$), allowing for both short-range ($H < 1/2$) and long-range ($H > 1/2$) dependencies. 

\par Fractional Brownian motion is a natural one-parameter extension of the classical Brownian motion with its statistical properties completely determined by the the Hurst parameter $H \in (0,1)$ \cite{biagini2008fbm}. 
\begin{definition}
    Let $H \in (0,1)$ be a constant. A fractional Brownian motion $\{B_t\}_{t\geq0}$ of Hurst parameter $H$ is a continuous and centered Gaussian process with covariance function
    \begin{equation}
        Cov(B_s,B_t) = \mathbb{E}[B_sB_t] = \frac{1}{2}(t^{2H} + s^{2H} - |t-s|^{2H}). 
        \end{equation}
\end{definition}

When $H = 1/2$, fractional Brownian motion reduces to standard Brownian motion with independent increments. For $H < 1/2$, it exhibits anti-persistent behavior where increments tend to reverse direction, resulting in rougher trajectories. On the other hand, $H > 1/2$ yields persistent behavior with smoother paths and positive long-range correlations.

\par However, when $H\neq \frac{1}{2}$, fractional Brownian motion $\{B_t\}_{t\geq 0}$ is not a semimartingale. Thus, we will not be able to use usual methods in It\^{o} calculus to manipulate fractional Brownian motion or related processes whenever $H\neq \frac{1}{2}$ \cite{biagini2008fbm}. To overcome that obstacle, in this work, we use a semimartingale approximation of fractional Brownian motion \cite{thao2006approximate}. 

\par Nuarlart has proposed fractional Brownian motion as 
\begin{equation}
    B_t = \int_0^{t}(t-s)^{\alpha} \di{W_s}
\end{equation}
where $\{W_t\}_{t\geq 0}$ is a standard Brownian motion and $\alpha=H-\frac{1}{2}$ \cite{nualart2006fbm}. The following theorem gives a semimartingale approximation of the fractional process $\{B_t\}_{t\geq 0}$.

\begin{theorem}
\label{thm:approximate fractional Brownian motion}
    Let $\{B_t\}_{t\geq 0}$ be a fractional Brownian motion with Hurst parameter $H$. For every $\varepsilon >0$ we define
    \begin{displaymath}
        B_t^{\varepsilon} = \int_{0}^{t} (t-s+\varepsilon)^\alpha \mathrm{d}W_s
    \end{displaymath}
    where $\alpha = H-\frac{1}{2}$.
    \begin{enumerate}[label=(\roman*)]
        \item  The process $\{B_t^{\varepsilon} : 0\leq t \leq T \}$ is a semi-martingale. Moreover, 
        \begin{displaymath}
            B_t^{\varepsilon} = \int_{0}^{t} \varphi(s) \mathrm{d}s + \varepsilon^{\alpha }W_t
        \end{displaymath}
        where $\varphi(t) = \int_{0}^{t} \alpha (t-u+\varepsilon)^{\alpha-1} \mathrm{d}W_u$.

        \item $B_t^{\varepsilon}$ converges to $B_t$ in $L^2-$sense when $\varepsilon$ tends to 0. This convergence is uniform with respect to $t \in [0,T]$.
    \end{enumerate}
\end{theorem}
This approximation introduces a parameter $\varepsilon$ and allows us to define a stochastic Schr\"{o}dinger equation with multiplicative noise driven by the approximated fractional Brownian motion.

\par The remainder of this article is organized as follows: Section \ref{sec: Stochastic Schrodinger Equation} develops the mathematical framework for incorporating noise driven by fractional Brownian motion into the Schr\"{o}dinger equation, resulting in a stochastic Schr\"{o}dinger equation with multiplicative noise. Section \ref{sec: Exact Cover - 3 problem} introduces the Exact Cover-3 problem, instance selection and its transformation into the Ising model. Section \ref{sec: Defining Hamiltonians} defines the initial and final Hamiltonians and describes the adiabatic evolution path. Section \ref{sec: Numerical Simulation} outlines our numerical simulation methodology. Section \ref{sec: Results and Discussion} presents the results of our simulations, comparing the performance of the quantum adiabatic algorithm with and without noise driven by fractional Brownian motion across various test cases.

\section{Stochastic Schr\"{o}dinger Equation}
\label{sec: Stochastic Schrodinger Equation}
Let $\mathcal{H}$ be an arbitrary Hilbert space. We begin with the generic form of linear stochastic differential equation for the non-normalized process $\ket{\psi(t)} \in \mathcal{H}$ with multiplicative noise induced by fractional Brownian motion $\{B_t^{\varepsilon} : t\geq 0\}$
\begin{align}
\label{eq:general linear schrodinger}
    \di{\ket{\psi(t)}} &= A(t)\ket{\psi(t)}\di{t} + B(t)\ket{\psi(t)}\di{B_t^{\varepsilon}} \nonumber \\
    &= A(t)\ket{\psi(t)}\di{t} + B(t)\ket{\psi(t)}\Big[  \varphi(t) dt + \varepsilon^{\alpha} dW_t\Big]
\end{align}
from which we derive
\begin{equation}
    \ket{\di{\psi(t)}} = \Big( A(t) + \varphi(t)B(t) \Big) \ket{\psi(t)}\di{t} + \varepsilon^{\alpha} B(t)\ket{\psi(t)}\di{W_t} \label{eq: substituted linear SE}
\end{equation}
where $A(t)$ and $B(t)$ are linear operators on $\mathcal{H}$. We closely follow the derivation of the stochastic Schr\"{o}dinger equation in \cite{barchielli2010stochastic}.
In this work, we consider $X_t = B_t^{\varepsilon}$ as presented in theorem \ref{thm:approximate fractional Brownian motion}, and hence \eqref{eq:general linear schrodinger} can be restated as

By taking the conjugate transpose of \eqref{eq: substituted linear SE} we have,
\begin{equation}
    \bra{\di{\psi(t)}} = \bra{\psi(t)}\Big(A^{\dagger}(t) + \varphi(t)B^{\dagger}(t)\Big)\di{t} + \varepsilon^{\alpha}\bra{\psi(t)}B^{\dagger}(t)\di{W_t}
\end{equation}

Due to the normalization factor, that is, $||\psi(t)||^2 = 1$, the stochastic differential 
\begin{equation}
    \di{||\psi(t)||^2} = 0.
    \label{eq: stoch differential zero}
\end{equation}
Now, by definition of the norm, $\di{||\psi(t)||^2} = \di{\langle\psi(t)|\psi(t)\rangle}$ and expanding using Ito's lemma \cite{oksendal2010sdes}, we have
\begin{align}
    \di{||\psi(t)||^2} &= \langle\di{\psi(t)}|\psi(t)\rangle + \langle\psi(t)|\di{\psi(t)}\rangle + \langle\di{\psi(t)}|\di{\psi(t)}\rangle \nonumber\\
    &=\bra{\psi(t)}\Big(A^{\dagger}(t) + \varphi(t)B^{\dagger}(t) + A(t) + \varphi(t)B(t) \nonumber \\ 
    & \quad\quad + \varepsilon^{2\alpha}B^{\dagger}(t)B(t)\Big)\ket{\psi(t)}\di{t} + \bra{\psi(t)}\Big(\varepsilon^{\alpha}(B^{\dagger}(t) + B(t))\Big)\ket{\psi(t)}\di{W_t} \nonumber
\end{align}
In order to ensure \eqref{eq: stoch differential zero},
\begin{align}
    A(t) + A^{\dagger}(t) + \varphi(t)(B^{\dagger}(t) + B(t)) + \varepsilon^{2\alpha} B^{\dagger}(t)B(t) = 0 \label{eq: dt=0}\\
    \varepsilon^{\alpha}\Big(B^{\dagger}(t) + B(t)\Big) = 0 \label{eq: dW=0}
\end{align}
Equations \eqref{eq: dt=0},\eqref{eq: dW=0} implies that there are two self adjoint operators $H(t)$ and $S(t)$ such that $B(t) = -\imath S(t)$ and $A(t) = -\imath H(t) - \frac{1}{2}\varepsilon^{2\alpha}S^2(t)$. Here, $H(t)$ is called the $\mathit{effective}$ Hamiltonian of the system \cite{barchielli2010stochastic}. Next, we rewrite \eqref{eq: substituted linear SE} by substituting $A(t)$ and $B(t)$ as,
\begin{equation}
    \di{\ket{\psi(t)}} = \Big(-\imath H(t) - \frac{1}{2}\varepsilon^{2\alpha}S^2(t) - \imath \varphi(t)S(t)\Big)\ket{\psi(t)}\di{t} - \imath \varepsilon^{\alpha}S(t)\ket{\psi(t)}\di{W_t} \label{eq: with S(t)}
\end{equation}
We focus on the specific case where $S(t)$ is the effective Hamiltonian $H(t)$. As we demonstrate below, this choice results in multiplicative noise proportional to the effective Hamiltonian $H(t)$. Then \eqref{eq: with S(t)} reduces to the stochastic Schr\"{o}dinger equation
\begin{equation}
    \di{\ket{\psi(t)}} = \Big(-\imath H(t) - \frac{1}{2}\varepsilon^{2\alpha}H^2(t) - \imath \varphi(t)H(t)\Big)\ket{\psi(t)}\di{t} - \imath \varepsilon^{\alpha}H(t)\ket{\psi(t)}\di{W_t} \label{eq: SSE with approx fractional Brownian motion noise}
\end{equation}
which will govern the time evolution of quantum states subjected to multiplicative noise driven by fractional Brownian motion.
\par There are several key aspects of this equation that merit consideration. Firstly, the formal solution of \eqref{eq: SSE with approx fractional Brownian motion noise} can be expressed as
\begin{equation*}
   \ket{\psi(t)} = U(t,t_0)\ket{\psi(t_0)}
\end{equation*}
where $U(t,t_0)$ is the time evolution operator given by
\begin{equation}
    U(t,t_0) = \mathcal{T}exp\;\bigg\{-\imath \int_0^t H(s) + \varphi(s)H(s)\di{s} - \imath \varepsilon^{\alpha}\int_0^t H(s)\di{W_s}\bigg\}
    \label{eq:formal solution of SSE}
\end{equation}
and $\ket{\psi(t_0)}$ is the state vector at time $t_0$ which provides the initial conditions for the time evolution of $\ket{\psi(t)}$ \cite{barchielli2010stochastic}. $\mathcal{T}exp$ denotes the time ordered exponential which is a generalization of the ordinary operator exponential. The need of time ordering arises due to the fact that time dependent operators $H(t_1)$ and $H(t_2)$ does not necessarily commute whenever $t_1\neq t_2$.
Most importantly, we can view \eqref{eq:formal solution of SSE} as time evolution being generated by the time-dependent Hamiltonian
\begin{align}
    \hat{H}(t) &= H(t) + \varphi(t)H(t) + \varepsilon^{\alpha}\xi(t)H(t) \nonumber \\
    &= H(t) + H(t)\underbrace{(\varphi(t) + \varepsilon^{\alpha}\xi(t))}_{noise}
\end{align}
where noise is integrated proportional to the original time dependent Hamiltonian. Here, $\xi(t)$ also known as white noise.

\section{Exact Cover - 3 Problem}
\label{sec: Exact Cover - 3 problem}

In order to test our hypothesis that introducing multiplicative noise driven by fractional Brownian motion proportional to the system Hamiltonian can enhance the performance
of the quantum adiabatic algorithm, we use the NP-complete Exact Cover 3 (EC3) problem (see \cite{farhi2001quantum,farhi2000numerical}). An $n$-bit instance of EC3 consist of $m$ distinct clauses $C_1, C_2, \dots, C_m$ connected by $\wedge$ operators,
\begin{equation}
    C_1 \wedge C_2 \wedge\dots\wedge C_m
\end{equation}
where each clause $C_l$ has exactly 3 distinct bits $(x_i, x_j, x_k)$ out of the n bits $x_1,x_2,\dots,x_n$. Each clause $C_l$ turns out true if the three bits satisfy the constraint $x_i+x_j+x_k=1$. In other words, one of the bits must be assigned 1 while the other two bits are assigned 0. Solving the Exact Cover 3 problem is finding if there exists a value assignment to the $n$ bits that simultaneously satisfy all the clauses. Furthermore, we can denote an $n$-bit instance of EC3 by a list of triples $(i,j,k)$ where $i,j,k \in \{1,2,\dots,n\}$ refers to the three distinct bits featured in each clause. For example, the following is an example of a 6-bit instance of EC3.
\begin{equation}
    [(1,3,6),(2,4,5),(3,5,6),(1,2,3),(3,4,6)]
\end{equation}
This instance consists of five clauses. If we observe carefully, it is easy to verify that assignment $(0,1,0,0,0,1)$ satisfies all five clauses. In fact, this is the only assignment that will satisfy all the clauses in this instance. However, determining the existence of at least one such satisfying requires an exhaustive brute-force search over all $2^6=64$ possible assignments. As the number of bits $n$ increases, the search space grows exponentially, making this approach computationally inefficient.

We have stated EC3 as a decision problem requiring a \textit{yes} or \textit{no} answer. However, quantum adiabatic algorithm is designed to solve optimization problems rather than decision problems. Hence, we consider an optimization version of EC3 which can be stated as a minimization problem to find the assignment that violates the minimum number of clauses. If the minimum number of clauses violated by this assignment is zero, then there exists at least one assignment satisfying all clauses. Conversely, if the number of violated clauses is positive, then there exists no such satisfying assignment.

The minimization problem can be defined \cite{mcgeoch2014adiabatic} as sum of individual clause functions $f_{C_l}$ as follows.
\begin{equation*}
    \min_{\mathbf{x}\in \{0,1\}^n}f(\mathbf{x})
\end{equation*}
\begin{equation*}
     \text{where} \qquad f_{C_l}(\mathbf{x}) = (1-x_i-x_j-x_k)^2 \hspace{1cm} ; \quad C_l = (x_i,x_j,x_k)
\end{equation*}
\begin{equation}
    \text{and}\qquad f(\mathbf{x}) = \sum_{l=1}^{m}f_{C_l}(\mathbf{x})
    \label{eq: Objective function}
\end{equation}
Observe that each individual clause function $f_{C_l}$ is nonnegative and equals to zero precisely when the clause is satisfied. i.e., whenever one bit is assigned 1 and the other two bits are 0. As a result, the objective function $f$ is also nonnegative and vanishes only when all clauses are satisfied. 

\par In accordance with the methodologies established by Farhi et al. \cite{farhi2000numerical,farhi2001quantum}, we have restricted our analysis to instances with a unique satisfying assignment. This constraint is motivated by the observation that the quantum adiabatic algorithm performs better for problems with multiple satisfying assignments.

We use the following algorithm to generate random instances of EC3. We start by randomly picking three distinct bits with uniform probability from the set of bits $\{1,2,\dots,n\}$, thereby creating an initial $n$-bit instance containing a single clause. Then we count the number of satisfying assignments. Next, we add another clause to this instance by again picking three distinct bits with uniform probability from the same set of $n$ bits. Following each addition, we recalculate the number of satisfying assignments.This iterative process continues until the instance becomes unsatisfiable (zero satisfying assignments) or uniquely satisfiable (exactly one  satisfying assignment). We retain only the uniquely satisfiable instances for our analysis, discarding and reinitiating the generation procedure for unsatisfiable instances. It is important to note that the number of clauses in the instance varies from instance to instance even though the number of bits is fixed at $n$.

\section{Defining Hamiltonians}
\label{sec: Defining Hamiltonians}

\subsection{Final (Problem) Hamiltonian}
\begin{figure}
    \centering
    \includegraphics[width=0.7\linewidth]{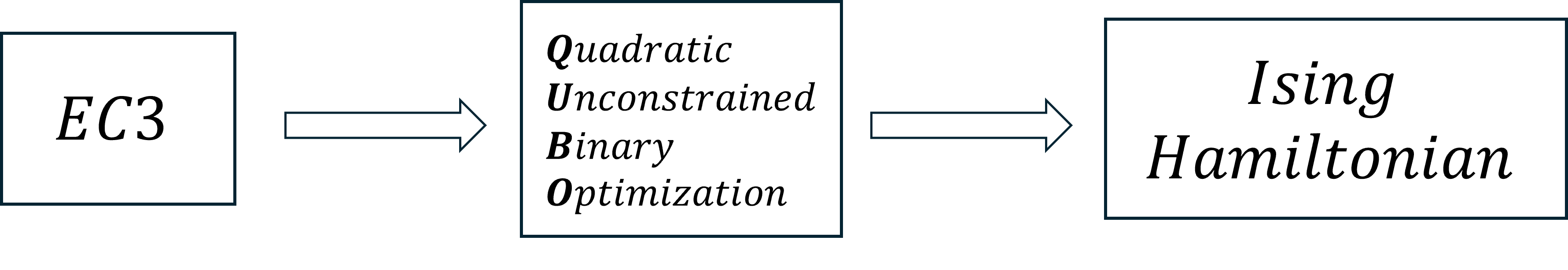}
    \caption{The process of defining the final Hamiltonian for the Exact cover 3 problem}
    \label{fig: Ising Hamiltonian process}
\end{figure}

In order to define the final Hamiltonian that encodes the solution of an EC3 problem in its ground state, we follow the process illustrated in Figure \ref{fig: Ising Hamiltonian process}. Initially, the binary objective function \eqref{eq: Objective function} is reshaped to take the form of a Quadratic Unconstrained Binary Optimization (QUBO) problem. This transformation is quite straight foward. Next, we use the QUBO to Ising transformation \cite{isingteachinganoldproblem} to arrive at the final Hamiltonian. This Hamiltonian will have the form
\begin{equation}
    \label{eq: H_F with comp basis}
    H_F = \sum_{\mathbf{x}\in \{0,1\}^n} h(\mathbf{x})\ket{\mathbf{x}}\bra{\mathbf{x}}
\end{equation}
where $h$ is equal to $f$ up to a constant.
Note that \eqref{eq: H_F with comp basis} implies $H_F$ is diagonal in the computational basis and hence the eigenstates will be the computational basis states. Due to the fact that each possible $n$-bit assignment $\mathbf{x}= (x_1,x_2,\dots, x_n) \in \{0,1\}^{n}$ has a corresponding computation basis state $\ket{\mathbf{x}} = \bigotimes_{i=n}^{1}\ket{x_i} =\ket{x_nx_{n-1}\dots x_1}$, each possible assignment corresponds to an eigenstate of $H_F$. Moreover, the ground state will correspond to the $n$-bit assignment that minimizes $h$.

Let us now focus on the first transformation; from objective function to QUBO. Starting from the expression in \eqref{eq: Objective function}, we multiply out and collect terms to arrive at
\begin{equation}
    f(\mathbf{x}) = m - \sum_{i=1}^{n}B_ix_i + \sum_{i<j}^{n}C_{ij}x_ix_j \hspace{3cm} (\,\because x_i^2 = x_i)
\end{equation}
where $B_i$ count the number of clauses that $x_i$ appears in while $C_{ij}$ counts the number of clauses in which $x_i$ and $x_j$ appear together. Now, let us define $g$ such that,
\begin{equation}
    g(\mathbf{x}) = - \sum_{i=1}^{n}B_ix_i + \sum_{i<j}^{n}C_{ij}x_ix_j 
    \label{eq: g(x)}
\end{equation}
Then $f$ and $g$ differ only by a constant term $m$. Since a constant term does not affect the minimization of a function, we choose to minimize $g$. Now, notice that $g(\mathbf{x)}$ in \eqref{eq: g(x)} is in the form of a QUBO:
\begin{equation}
    \sum_{i=1}^{n}Q_{ii}x_i + \sum_{i<j}^{n}Q_{ij}x_ix_j
\end{equation}
where $Q$ is the upper triangular matrix defined such that
\begin{equation}
    Q_{ij} = 
    \begin{cases}
        C_{ij} \quad & \text{if } \quad  i < j \\
        0 \quad & \text{if } \quad i > j\\
        -B_i \quad & \text{if } \quad i = j.
    \end{cases}
\end{equation}

Subsequently, we use the QUBO to Ising Hamiltonian transformation \cite{isingteachinganoldproblem} to generate the problem Hamiltonian. This Ising Hamiltonian will take the form
\begin{equation}
    H_F = \sum_{i=1}^{n}h_{i}\sigma_i^{z} + \sum_{i<j}^{n}J_{ij}\sigma_i^{z}\sigma_j^{z}
\end{equation}
where $h_i$ and $J_{ij}$ are derived form the entries of the QUBO matrix. $\sigma_i^{z}$ is defines as $\sigma_i^{z} = I\otimes I \otimes \dots \otimes \sigma^z \otimes \dots \otimes I$ ($\sigma^z$ is in the $i^{th}$ position) and $\sigma_i^{z}\sigma_j^{z}$ such that $\sigma^z$ is in both $i^{th}$ and $j^{th}$ positions while $\sigma^z$ is the Pauli Z matrix.

It is worth mentioning that in the process described in Figure \ref{fig: Ising Hamiltonian process}, an $n$-bit instance of EC3 is converted to a QUBO problem with $n \times n$ $Q$ matrix. Consequently, this will be transformed to the problem of finding the ground state of the Ising Hamiltonian whose dimensions are $2^n \times 2^n$. Thus, we will need $n$ qubits to map an $n$-bit instance of EC3 to solve using an adiabatic quantum computer.

\subsection{Initial Hamiltonian}
Let the initial Hamiltonian be
\begin{equation}
    H_I = -\sum_{i=1}^{n}\sigma_i^x
    \label{eq: initial Hamiltonian}
\end{equation}
which is diagonal in the Hadamard basis. A Hamiltonian that is diagonal in a basis that is orthogonal to the basis in which the final Hamiltonian $H_F$ is diagonal is employed as the initial Hamiltonian to minimize eigenvalue crossings during adiabatic evolution. Moreover, the ground state of $H_I$ is the uniform superposition of computational basis states 
\begin{equation}
    \ket{\psi(0)} =\frac{1}{\sqrt{2^n}}\sum_{\mathbf{x}\in\{0,1\}}\ket{\mathbf{x}}
\end{equation}
from which we will initialize the adiabatic time evolution.

\subsection{Adiabatic path}
We choose $s(t)=t/T$ where $T$ is the total evolution time. Then the time dependent Hamiltonian is defined as 
\begin{equation}
    H(t) = \bigg(1-\frac{t}{T}\bigg)H_I + \bigg(\frac{t}{T}\bigg)H_F
\end{equation}

\section{Numerical Simulation}
\label{sec: Numerical Simulation}

As discussed in section \ref{sec: Stochastic Schrodinger Equation}, the time evolution of the system is taken place according to \eqref{eq: SSE with approx fractional Brownian motion noise}, the stochastic Schr\"{o}dinger equation driven by approximate fractional Brownian motion. Further, we showed that there is no closed-form solution for \eqref{eq: SSE with approx fractional Brownian motion noise} and the formal solution \eqref{eq:formal solution of SSE} consists of a time-ordered exponential which is impractical to simulate due to its complexity. Thus, we are forced to use an approach that approximates the time evolution within a time step. This time step is chosen to be relatively small in order to simulate the physical system as accurately as possible. The convergence properties of this method in the noiseless case are established in \cite{van2001howpowerful}. For the stochastic generalization relevant to our work , analogous lemmas may be derived from the results in \cite{fernando2015stochastic}.

We begin by partitioning the interval $[0,T]$ in to $N$ uniform subintervals such that $0=t_0<t_1<\dots<t_N=1$. Consider such a time step $[t_i, t_{i+1}]\subset[0,T]$. Let $U(t_{i+1},t_i)$ be the time evolutionary operator acting on the Hilbert space.

Next we assume that the time dependent Hamiltonian $H(t)$ and the function $\varphi:[0,T]\longrightarrow \mathbb{R}$ defined in theorem \ref{thm:approximate fractional Brownian motion} are constant inside this time interval. Set $H=H(t_i)$ and $\varphi=\varphi(t_i)$. As a result, \eqref{eq: SSE with approx fractional Brownian motion noise} reduces to 
\begin{equation} 
    \label{eq: Discrete SSE with fractional Brownian motion noise}
    \di{\ket{\psi(t)}} = \bigg(-\imath H-\frac{1}{2}\varepsilon^{2\alpha}H^2 - \imath H\varphi \bigg)\ket{\psi(t)}\di{t} - \imath H\varepsilon^{\alpha}\ket{\psi(t)}\di{W_t}
\end{equation}
inside the time interval $[t_i, t_{i+1}]$.

Let $C = -\imath H -\imath H\varphi-\frac{1}{2}\varepsilon^{2\alpha}H^2$ and $D=-\imath H\varepsilon^{\alpha}$. Notice that $C$ and $D$ are constants and [C,D]=0. Then the vector-valued stochastic differential equation \eqref{eq: Discrete SSE with fractional Brownian motion noise} has the explicit solution \cite{kloeden1992numericalSDE} given by
\begin{equation*}
    \ket{\psi(t_{i+1})} = U(t_{i+1}, t_i)\ket{\psi(t_i)}
\end{equation*}
where 
\begin{align}
    U(t_{i+1}, t_i) &= \exp\Bigg\{\bigg(-\imath H-\imath H\varphi\bigg)\big(t_{i+1}-t_i\big)-\imath H\varepsilon^{\alpha}\Big(W(t_{i+1})-W(t_i)\Big)\Bigg\}
    \label{eq: U(t_i+1,t_i)}
\end{align}
Notice that here $W(t_{i+1})-W(t_i)$ is an increment of Brownian motion and hence $W(t_{i+1})-W(t_i) \sim N(0, t_{i+1}-t_i)$. In simulation, we can replace $W(t_{i+1})-W(t_i)$ with $Z_i\sqrt{ t_{i+1}-t_i}$ where $Z_i$ is a sequence of standard normal random variables. As a consequence \eqref{eq: U(t_i+1,t_i)} becomes,
\begin{equation}
    U(t_{i+1}, t_i) = \exp\Bigg\{\bigg(-\imath H-\imath H\varphi\bigg)\Delta t -\imath H\varepsilon^{\alpha}Z_i\sqrt{\Delta t}\Bigg\}
\end{equation}
where $\Delta t = \frac{T}{N}$.

Moreover, $\varphi$ in \eqref{eq: U(t_i+1,t_i)} can be simulated as discussed in \cite{thao2003fractal} in the following manner.

\begin{align}
    \varphi = \varphi(t_i) &= \alpha\int_0^{t_i} (t-s+\varepsilon)^{\alpha-1} \di{W_s} \nonumber \\
    & \approx \alpha\ \sum_{k=0}^{M-1}(t-k\frac{t}{M}+\varepsilon)^{\alpha-1}[W((k+1)\frac{t}{M})-W(k\frac{t}{M})]
\end{align}
where $M$ is the number of equal sub intervals in a partition of $[0,t_i]$. Similar to the previous case, $W((k+1)\frac{t}{M})-W(k\frac{t}{M}) \sim N(0,\frac{t}{M})$. Hence, in simulation, we replace by $g_k\sqrt{\frac{t}{M}}$ where $g_k$ is a sequence of independent standard normal random variables. Hence it follows that,
\begin{equation}
    \varphi = \alpha \sqrt{\frac{t}{M}} \sum_{k=0}^{M-1}(t-k\frac{t}{M}+\varepsilon)^{\alpha-1} g_k
\end{equation}

Finally, combining all the results in this section, we write the time evolved state $\ket{\psi(T)}$ such that
\begin{align}
    \ket{\psi(T)} &= U(T=t_N, t_{N-1})U(t_{N-1},t_{N-2})\dots U(t_2,t_1) \ket{\psi(0)} \\
    &= U(T,0)\ket{\psi(0)} \nonumber
\end{align}

The simulation of the quantum adiabatic algorithm in the presence of noise driven by fractional Brownian motion is summarized by Algorithm \ref{alg: QA alg with noise}. For comparison purposes, Algorithm \ref{alg: standard QA algorithm} presents the simulation of the standard quantum adiabatic algorithm without noise.

\begin{algorithm}[h]
\caption{Simulating quantum adiabatic algorithm in the presence of fractional Brownian motion-driven noise}
    \begin{algorithmic}[1]
    \Require Initial Hamiltonian $H_I$, Final Hamiltonian $H_F$, Evolution time $T$, Time step $\Delta t $, Initial state $\ket{\psi_0}$, Function $\varphi(t,\alpha,\varepsilon)$ $\alpha$ parameter, and $\varepsilon >0$. \newline
    \State $t \gets 0$ \Comment{Define start of evolution at time 0} \vspace{0.5mm}
    \State $\ket{\psi} \gets \ket{\psi_0}$ \Comment{Setting the initial state} \vspace{0.5mm}
    \While{$t < T$} \Comment{Algorithm terminates when $t = T$} \vspace{0.5mm}
        \State $H \gets (1 - \frac{t}{T}) H_I+ \frac{t}{T} H_F$  \Comment{Time dependent Hamiltonian} \vspace{0.5mm}
        \State $Z \gets $ Normal(0,1) \Comment{Sampling from standard normal distribution}\vspace{0.5mm}
        \State $ dW \gets $ $Z \sqrt{\Delta t}$ \vspace{0.5mm}
        \State $U \gets e^{-\imath H (1+\varphi(t,\alpha,\varepsilon)) \Delta t - \imath \varepsilon^{\alpha}HdW}$ \vspace{0.5mm}\Comment{Operator exponentiation}
            \State $\ket{\psi} \gets U \ket{\psi}$ \vspace{0.5mm}\Comment{Apply time evolution operator to state vector}
        \State $t \gets t + \Delta t$ \vspace{0.5mm}
    \EndWhile \vspace{0.5mm}
    \State \Return $\ket{\psi}$ \vspace{0.5mm}
    \end{algorithmic}
    \label{alg: QA alg with noise}
\end{algorithm}

\begin{algorithm}[h]
\caption{Simulating standard quantum adiabatic algorithm}
    \begin{algorithmic}[1]
    \Require Initial Hamiltonian $H_I$, Final Hamiltonian $H_F$, Evolution time $T$, Time step $\Delta t $, Initial state $\ket{\psi_0}$ \newline
    \State $t \gets 0$ \Comment{Define start of evolution at time 0} \vspace{0.5mm}
    \State $\ket{\psi} \gets \ket{\psi_0}$ \Comment{Setting the initial state} \vspace{0.5mm}
    \While{$t < T$} \Comment{Algorithm terminates when $t = T$} \vspace{0.5mm}
        \State $H \gets (1 - \frac{t}{T}) H_I+ \frac{t}{T} H_F$ \vspace{0.5mm} \Comment{Time dependent Hamiltonian}
        \State $U \gets e^{-\imath H \Delta t}$ \vspace{0.5mm}\Comment{Operator exponentiation} \vspace{0.5mm}
        \State $\ket{\psi} \gets U \ket{\psi}$ \vspace{0.5mm}\Comment{Apply time evolution operator to state vector}
        \State $t \gets t + \Delta t$ \vspace{0.5mm}
    \EndWhile \vspace{0.5mm}
    \State \Return $\ket{\psi}$ \vspace{0.5mm}
    \end{algorithmic}
    \label{alg: standard QA algorithm}
\end{algorithm}

In addition to the time evolved state, our analysis requires the ground state of the final Hamiltonian $H_F$ for calculations discussed subsequently. We determine this ground state by identifying the minimum value an the position of the minimum value in the diagonal entries of $H_F$.  This position corresponds to the ground state of $H_F$ in the computational basis.

To measure the performance of the algorithm, we employ two key metrics. We calculate the fidelity  $F$ between the ground state of the final Hamiltonian (solution state) and the time evolved state $\ket{\psi(T)}$ given by
\begin{equation}
    F = \big|\langle\psi(T)|\phi_g\rangle\big|^2
\end{equation}
where $\ket{\phi_g}$ is the ground state of the final Hamiltonian. In other words, fidelity is the probability that the actual measurement of $\ket{\psi(T)}$ will provide $\ket{\phi_g}$. In our case $\ket{\phi_g}$ is a computational basis state. Since we have already calculated the $2^n$ complex coefficients of $\ket{\psi(T)}$, we can simply read off the coefficient corresponding to the position of 1 in $\ket{\phi_g}$ and taking the absolute square of that coefficient will give $F$. Moreover, $\overline{F}$ will represent $F$ averaged over realizations.

We define Speedup (SP) similarly to \cite{cao2021speedup} as the ratio between the average fidelity with obtained with the quantum adiabatic algorithm in the presence of noise driven by fractional Brownian motion ($\overline{F}_{noise}$) and the fidelity with the standard quantum adiabatic algorithm ($F_0$),
\begin{equation}
    SP = \frac{\overline{F}_{noise}}{F_0}
\end{equation}

\section{Results}
\label{sec: Results and Discussion}

In this section, we present the results from numerical simulations of the quantum adiabatic algorithm applied to the NP-Complete Exact Cover 3 (EC3) problem. We perform our analysis on five avenues, and in each avenue, we compare the standard quantum adiabatic algorithm with the quantum adiabatic algorithm in the presence of noise driven by fractional Brownian motion.

\subsection{Performance on a random instance and a minimally \\ gapped instance}

\begin{figure}[h]
    \centering
    \begin{subfigure}{0.45\textwidth}
        \includegraphics[width=1\linewidth]{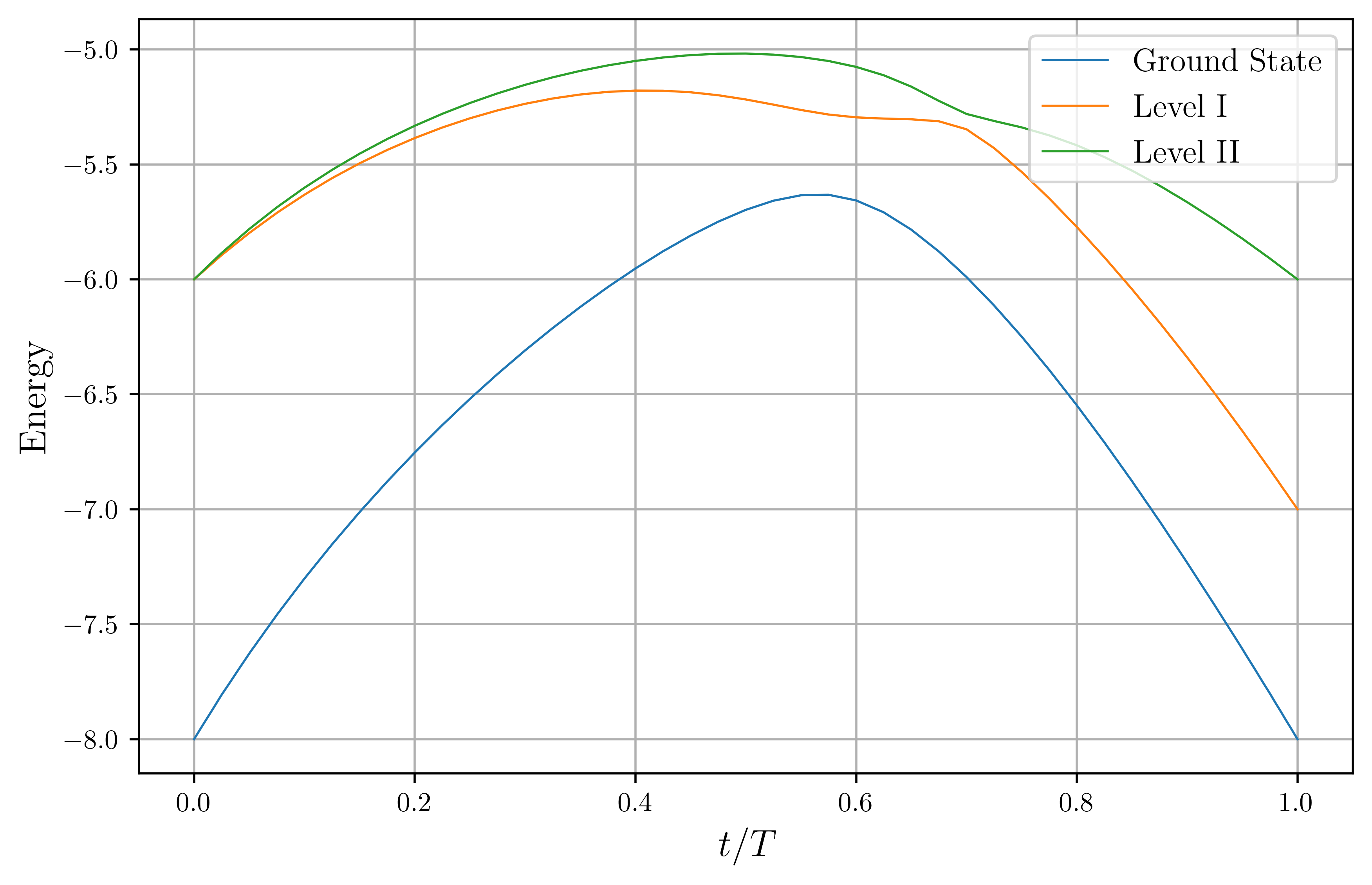}
        \caption{}
        \label{fig: energy random}
    \end{subfigure}
    \hspace{0.5cm}
    \begin{subfigure}{0.45\textwidth}
        \includegraphics[width=1\linewidth]{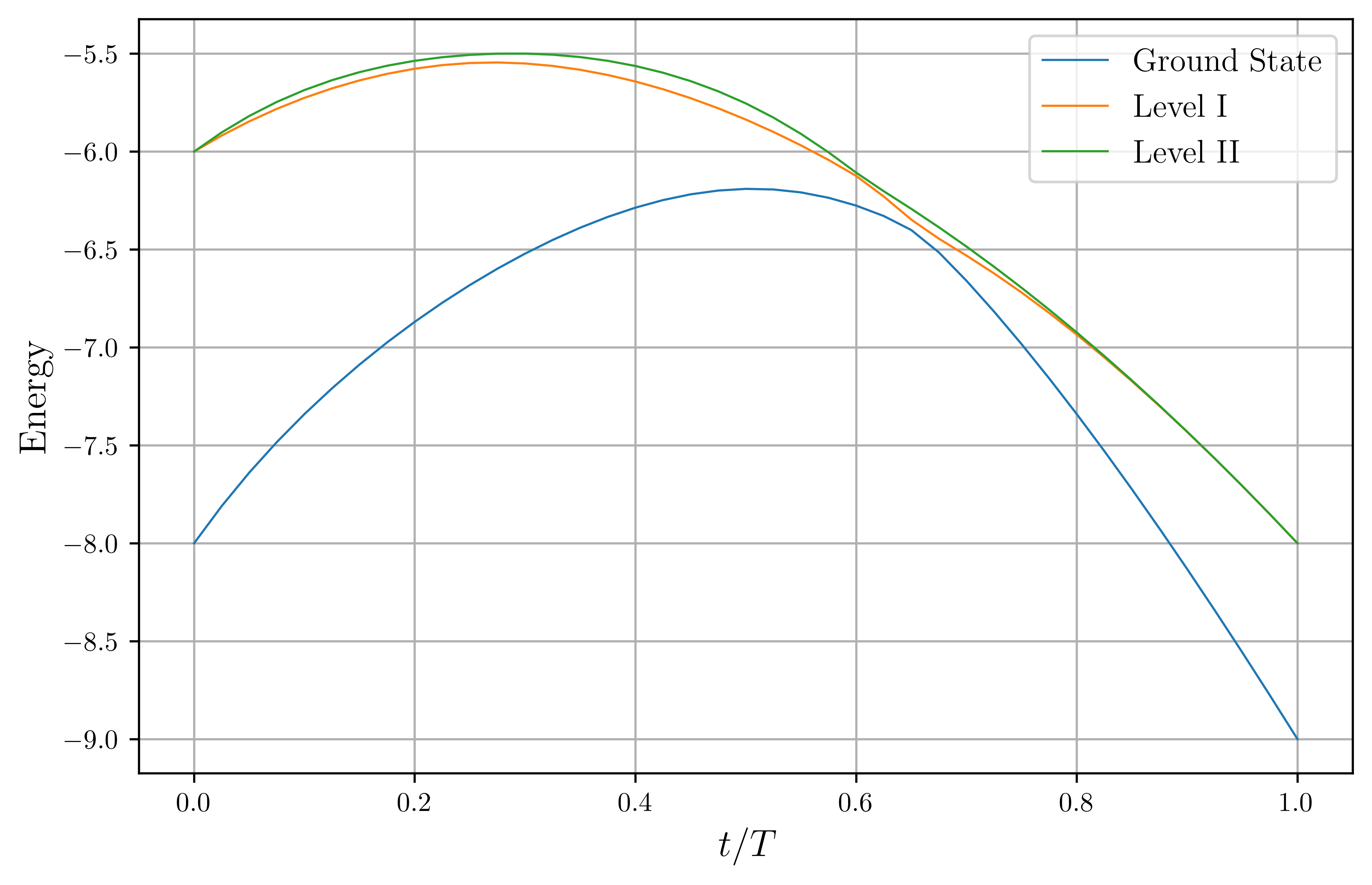}
        \caption{}
        \label{fig: energy hard}
    \end{subfigure}
    \caption{Energy diagrams representing the three lowest energy levels of the time dependent Hamiltonian $H(t)$ corresponding to the \textit{random} (a) and the \textit{hard} (b) instances of EC3. The time interval $[0,T]$ is linearly scaled to $[0,1]$. }
    \label{fig: energy both instances}
\end{figure}

Initially, we generate 10,000 instances of 8-bit EC3 problem according to the routine specified in Section \ref{sec: Exact Cover - 3 problem}. From that we select one instance randomly and another such that it has the lowest $\Delta_{min}$ considering their time dependent Hamiltonians without noise. We call the former the \textit{random} instance and the latter the \textit{hard} instance as the time $T$ to obtain a fixed fidelity is inversely proportional to $\Delta_{min}^2$ where $\Delta_{min}$ is the minimum energy gap between the instantaneous ground state and the instantaneous first excited state of the time dependent Hamiltonian $H(t)$ \cite{van2001howpowerful}. Figure \ref{fig: energy both instances} illustrates how the three lowest energy levels of $H(t)$ vary with respect to the scaled evolution time $t/T$ where $T$ is the evolution time. In both instances, we can observe that $\Delta_{min}$ occurs around 0.6-0.65 in scaled time. $\Delta_{min}$ of the random instance is 0.3613, while $\Delta_{min}$ of the hard instance is 0.0532. The time to achieve a fixed fidelity of 0.25 which is sufficiently large given the $2^8$ basis (solution) states, using the standard quantum adiabatic algorithm is $T=8$ for the random instance, whereas for the hard instance it is $T=200$.

\begin{figure}[p!]
    \centering
    \begin{subfigure}{0.8\textwidth}
        \includegraphics[width=\linewidth]{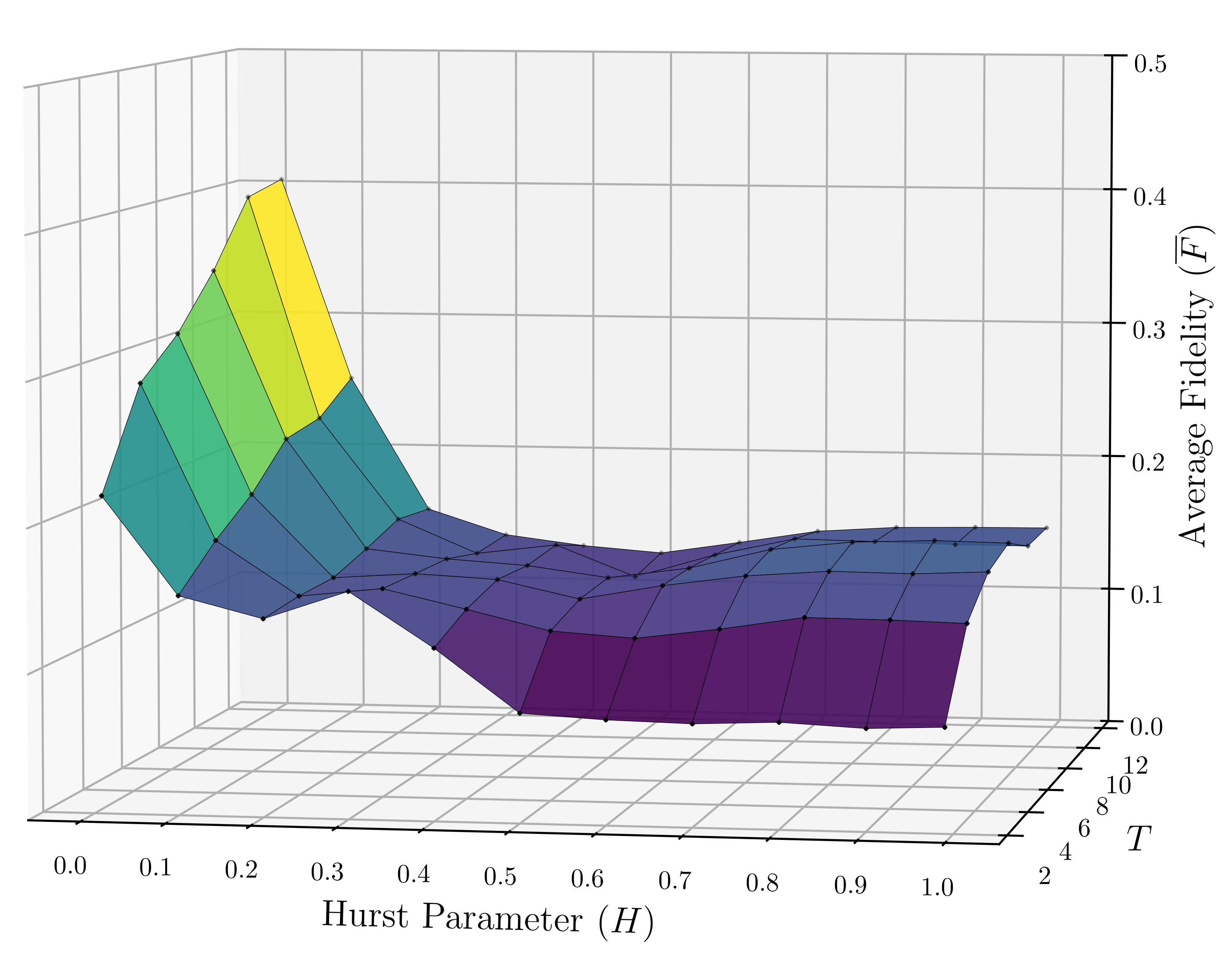}
        \caption{}
        \label{fig: 3Dplot random}
    \end{subfigure}
    \hspace{0.5cm}
    \begin{subfigure}{0.8\textwidth}
        \includegraphics[width=\linewidth]{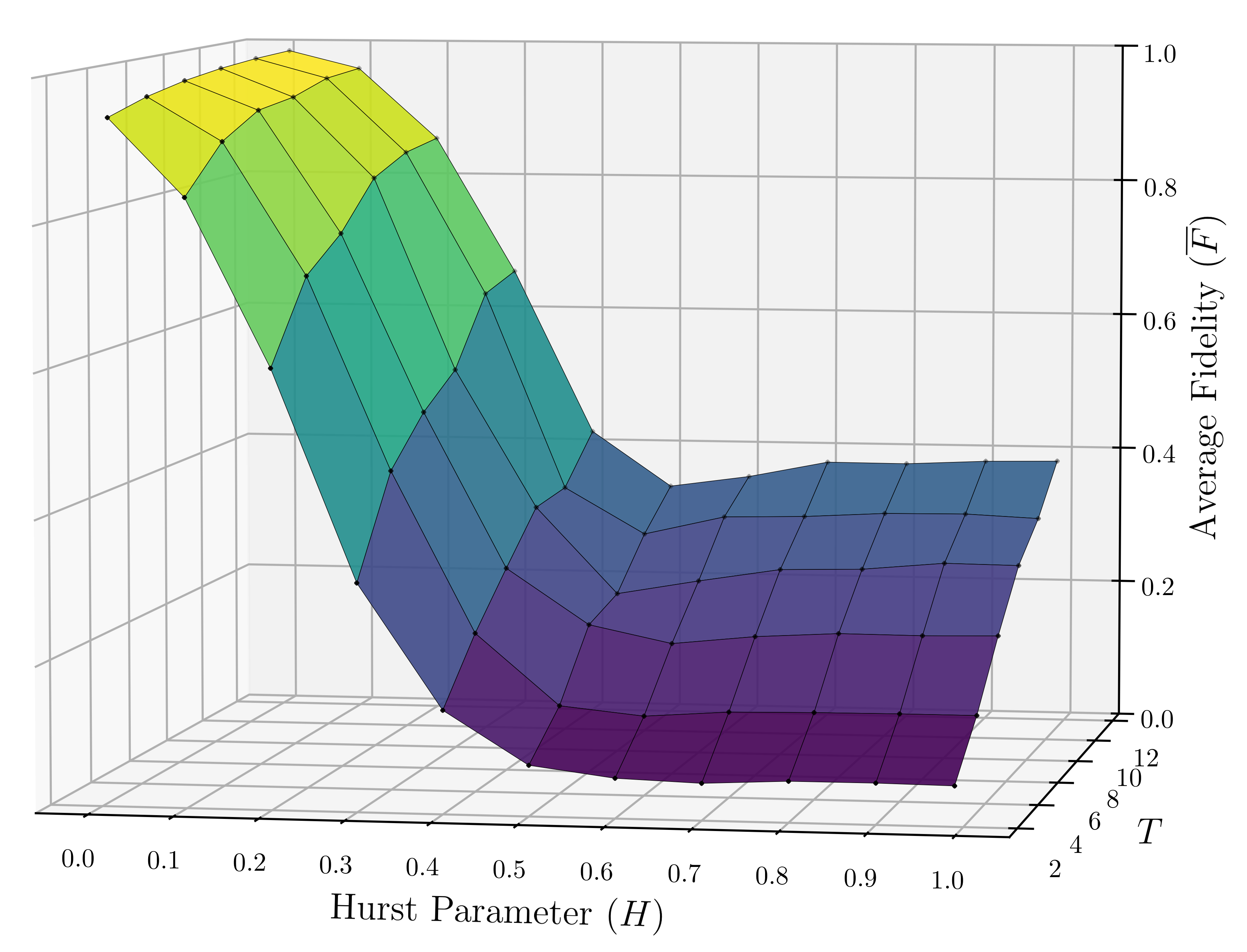}
        \caption{}
        \label{fig: 3Dplot hard}
    \end{subfigure}
    \caption{Average fidelity of the 8-bit \textit{random} (\ref{fig: 3Dplot random}) and \textit{hard} (\ref{fig: 3Dplot hard}) instances of EC3 plotted with different evolution times and Hurst parameter values. Fidelities $F$ of 100 runs are averaged to obtain $\overline{F}$ with time step $\Delta t=0.01$ and $\varepsilon = 10^{-3}$}
    \label{fig: 3Dplot both instances}
\end{figure}

\begin{figure}[p]
    \centering
    \begin{subfigure}{0.47\textwidth}
        \includegraphics[width=\linewidth]{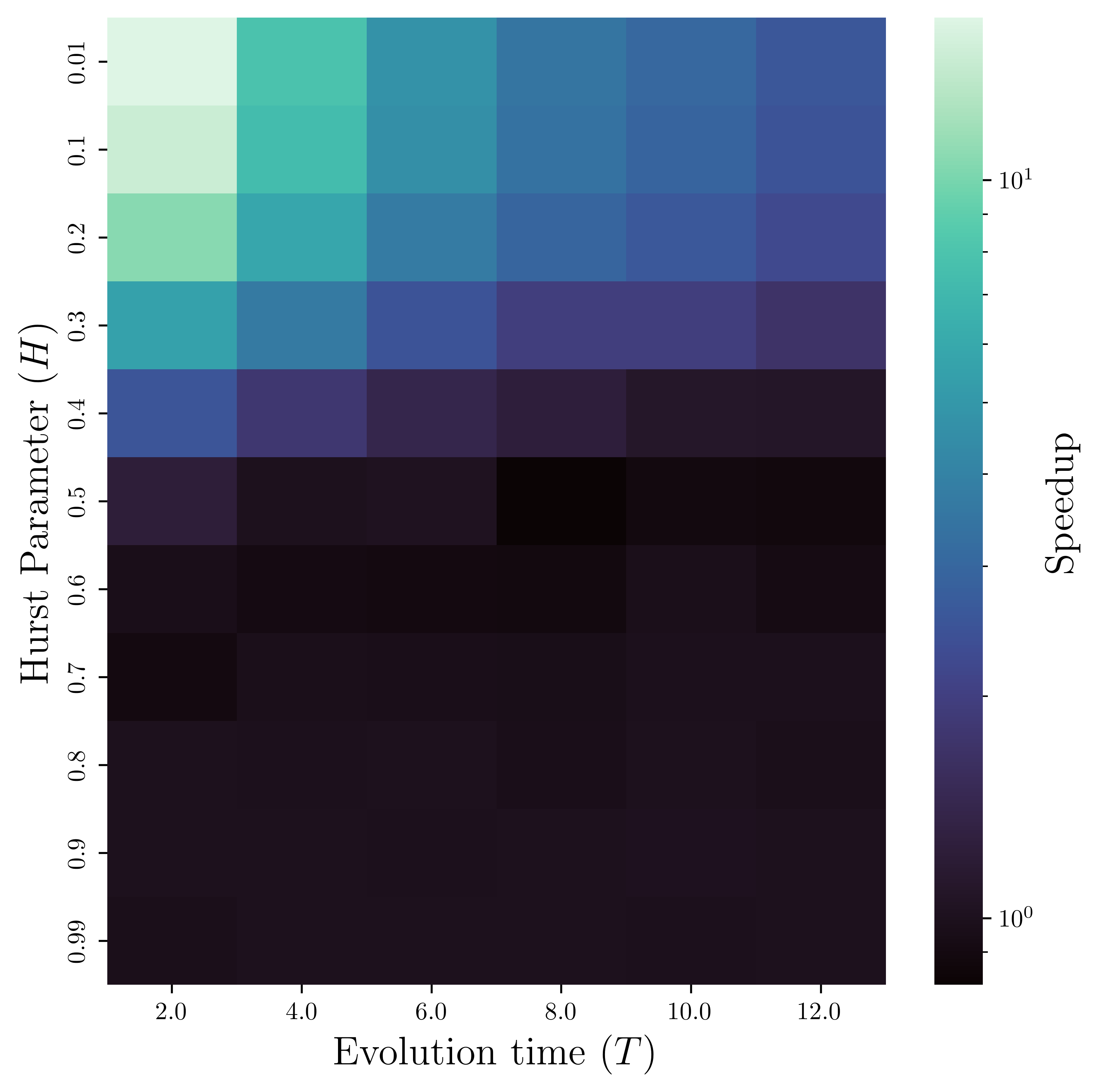}
        \caption{}
    \end{subfigure}
    \hspace{0.5cm}
    \begin{subfigure}{0.47\textwidth}
        \includegraphics[width=\linewidth]{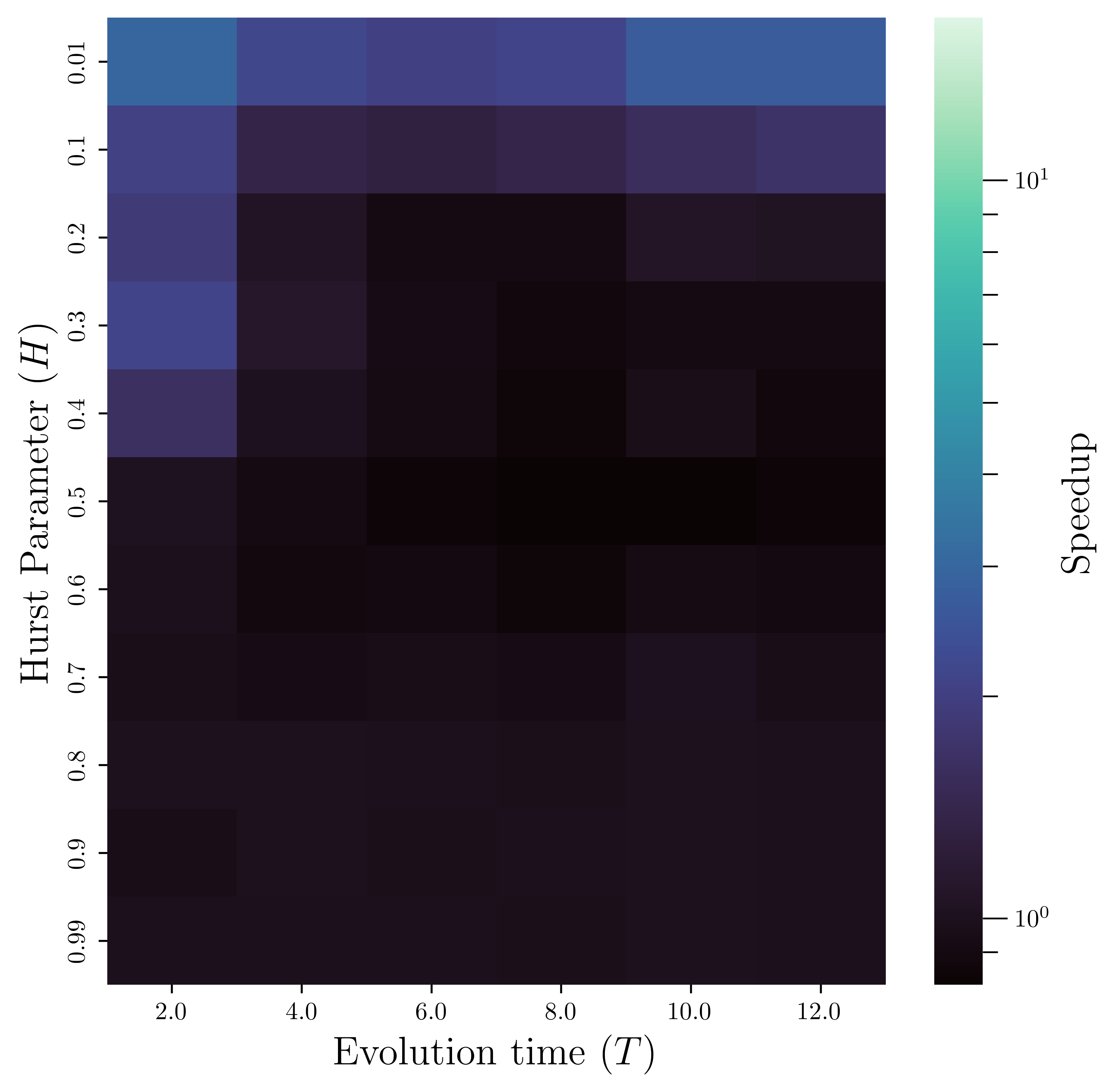}
        \caption{}
    \end{subfigure}
    \caption{Speedup achieved by the quantum adiabatic algorithm in the presence of noise driven by fractional Brownian motion compared to the standard quantum adiabatic algorithm of the random instance (a) and hard instance (b). }
    \label{fig: Heatmaps}
\end{figure}

We numerically simulate the quantum adiabatic algorithm with noise driven by fractional Brownian motion for the random and the hard instances with variations in evolution time ($T$) and Hurst parameter ($H$) and observe the behavior of average fidelity ($\overline{F}$). Furthermore, we simulate the standard quantum adiabatic algorithm for both instances as a reference and calculate the speedup (SP). The results are presented in Figure \ref{fig: 3Dplot both instances}. We can observe that in the presence of noise driven by fractional Brownian motion with Hurst parameter $0<H<\frac{1}{2}$ the average fidelity $\overline{F}$  increases in both instances over all evolution times. Even though the random instance demonstrates larger gains in average fidelity than the hard instance, it is important to note that both instances show sensitiveness to the noise, positively. This is even more clearly illustrated in Figure \ref{fig: Heatmaps} where we illustrate the speedups gained by the two instances. We note that the random instance records speedups up to 16 while the hard instance gain speedups up to 3.75.

It is evident that $\overline{F}$ increases as $H$ decreases from 0.5 to 0 at all evolution times in Figure \ref{fig: 3Dplot both instances} and reaches its maximum as $H$ approaches $0$. In figure \ref{fig: 3Dplot both instances} the edge on the left represents $H=0.01$ where the maximum average fidelities are recorded in our simulations. 

As discussed in Section \ref{sec: Introduction}, the regime of $0<H<\frac{1}{2}$ exhibits short-range memory and antipersistent behavior, resulting in increasingly rough sample paths. This can be viewed as an increase in noise strength bearing some similarities to the noise strength analyzed in the Markovian (white noise) case in \cite{xu2018adiabatic}. In that work, white noise, corresponding to $H=\frac{1}{2}$ in our model, was used with a tunable noise strength parameter, and it was shown that increasing this parameter improved average fidelities. However, since noise driven by fractional Brownian motion with $0<H<\frac{1}{2}$ is non-Markovian, our work extends this result to the non-Markovian regime, suggesting that beneficial noise processes are not limited to Markovian processes.

In the regime $\frac{1}{2} \leq H < 1$, we observe no improvement in average fidelity. However, importantly, since the speedup remains close to 1 in this regime, there is no decrease in average fidelity with respect to that achieved by the standard quantum adiabatic algorithm.

\begin{figure}[p]
    \centering
    \begin{subfigure}{0.45\textwidth}
        \centering
        \includegraphics[width=\linewidth]{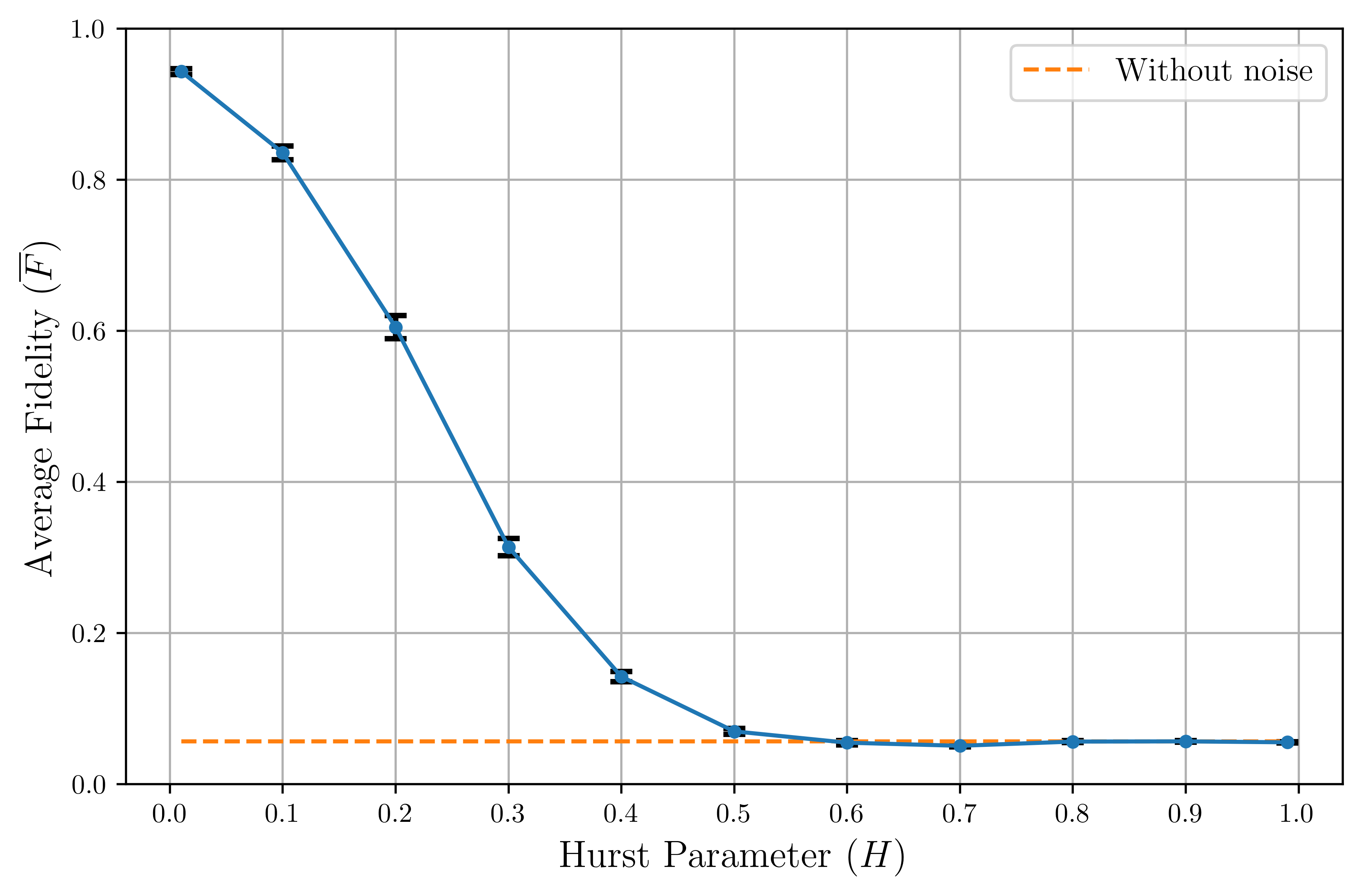}
        \caption{}
    \end{subfigure}
    \hfill
    \begin{subfigure}{0.45\textwidth}
        \centering
        \includegraphics[width=\linewidth]{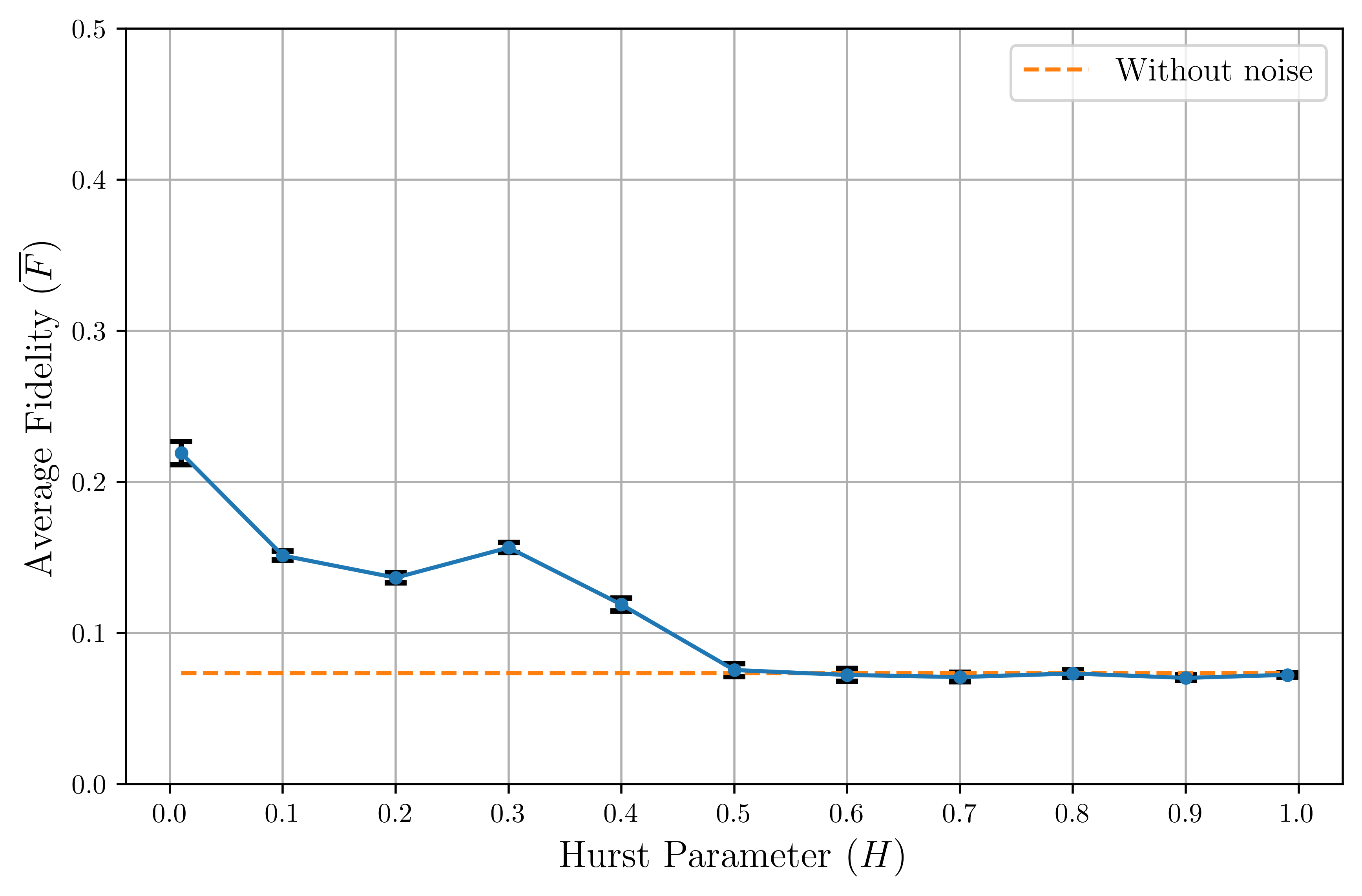}
        \caption{}
    \end{subfigure}

    \vskip 0.5cm 
    \begin{subfigure}{0.45\textwidth}
        \centering
        \includegraphics[width=\linewidth]{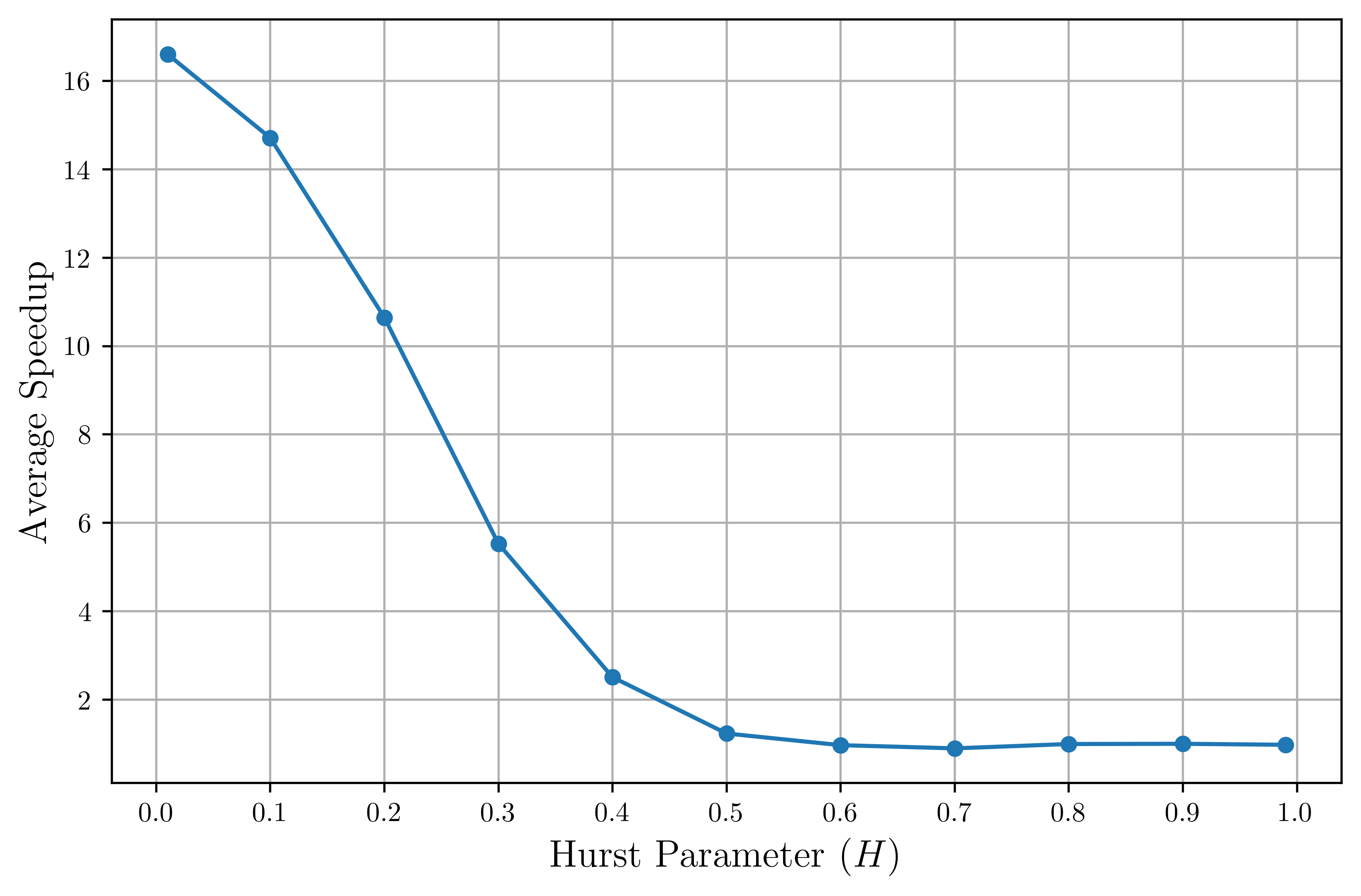}
        \caption{}
    \end{subfigure}
    \hfill
    \begin{subfigure}{0.45\textwidth}
        \centering
        \includegraphics[width=\linewidth]{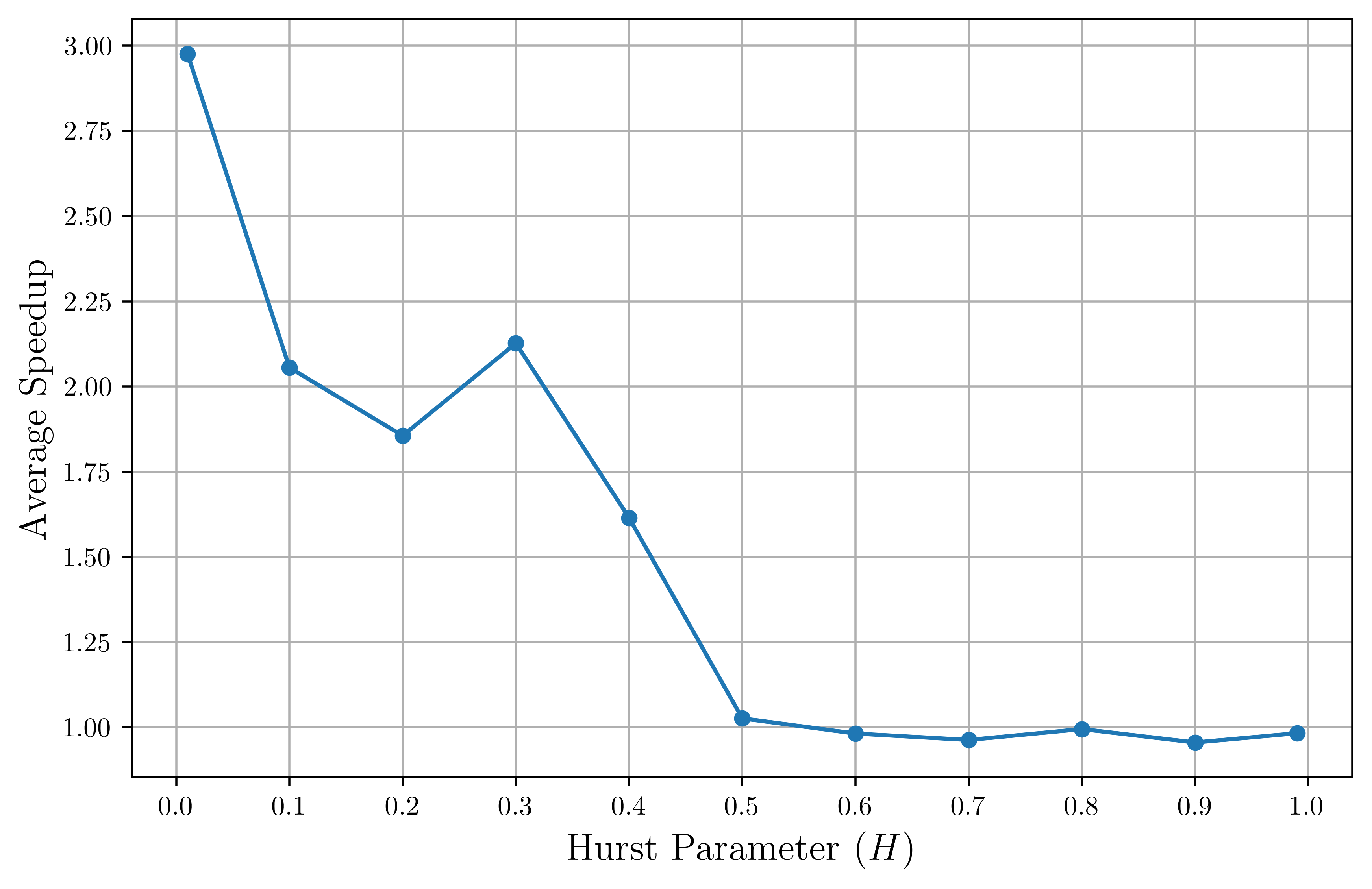}
        \caption{}
    \end{subfigure}
    \caption{Vertical cuts from the plot in Figure \ref{fig: 3Dplot both instances} at $T=4$ ((a) and (b)) and speedup plots at $T=4$ ((c) and (d)) with statistical error bars. Subfigures (a) and (c) correspond to the random instance while (b) and (d) correspond to the hard instance. The dotted orange lines in (a) and (b) represent the fidelity achieved by the standard quantum adiabatic algorithm.}
    \label{fig: vertical cuts}
\end{figure}

If one is interested in the increase in the speed of the algorithm in terms of the time taken to achieve a prescribed fidelity, for example 0.25, one can take a horizontal cut from the plots in Figure \ref{fig: 3Dplot both instances} at $\overline{F}=0.25$. However, we are interested in the increase of $\overline{F}$ with respect to the Hurst parameter at a fixed and limited evolution time. Thus, we take vertical cuts at $T=4$ and they are presented in Figure \ref{fig: vertical cuts}. These results demonstrate that for limited evolution times, we can achieve similar, if not better, average fidelities by performing the quantum adiabatic algorithm in the presence of noise driven by fractional Brownian motion with Hurst parameter $0<H<\frac{1}{2}$ than running for longer evolution times with the standard quantum adiabatic algorithm.

\subsection{A 4-bit example for probability amplitudes}

\begin{figure}
    \centering
    \begin{subfigure}{0.85\textwidth}
        \includegraphics[width=\linewidth]{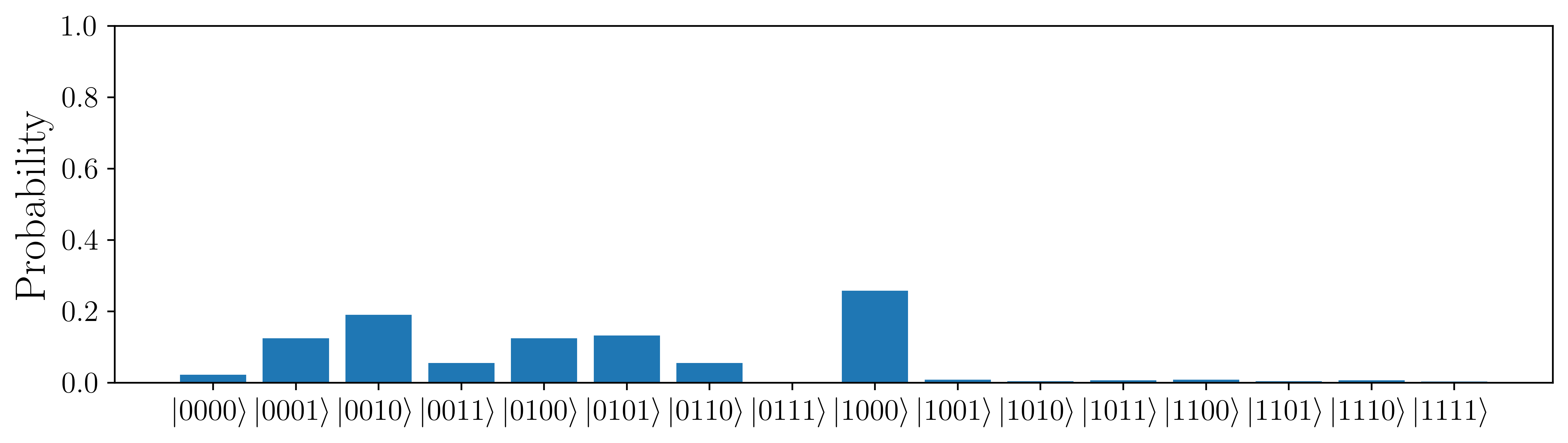}
    \caption{without noise}
    \label{}
    \end{subfigure}

    \begin{subfigure}{0.85\textwidth}
        \includegraphics[width=\linewidth]{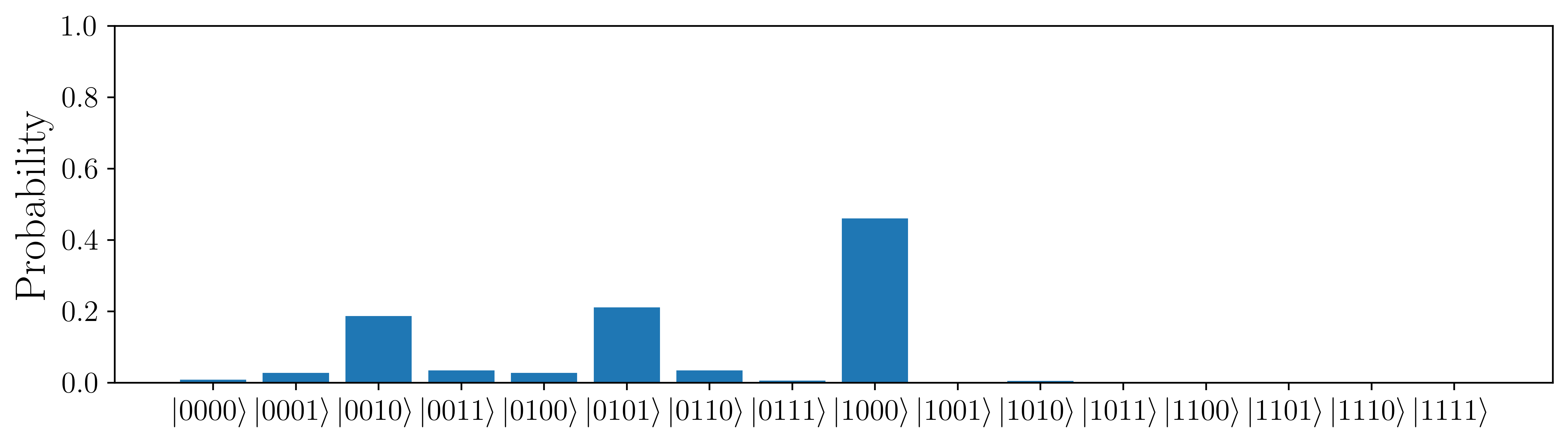}
    \caption{$H=0.3$}
    \label{}
    \end{subfigure}

    \begin{subfigure}{0.85\textwidth}
        \includegraphics[width=\linewidth]{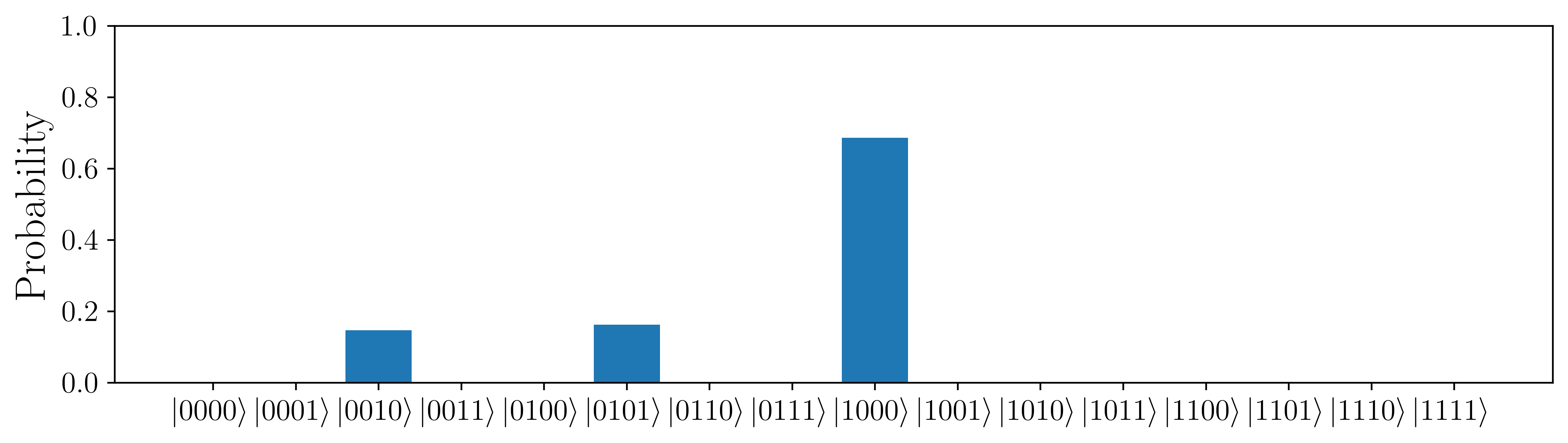}
    \caption{$H=0.2$}
    \label{}
    \end{subfigure}
    
    \begin{subfigure}{0.85\textwidth}
        \includegraphics[width=\linewidth]{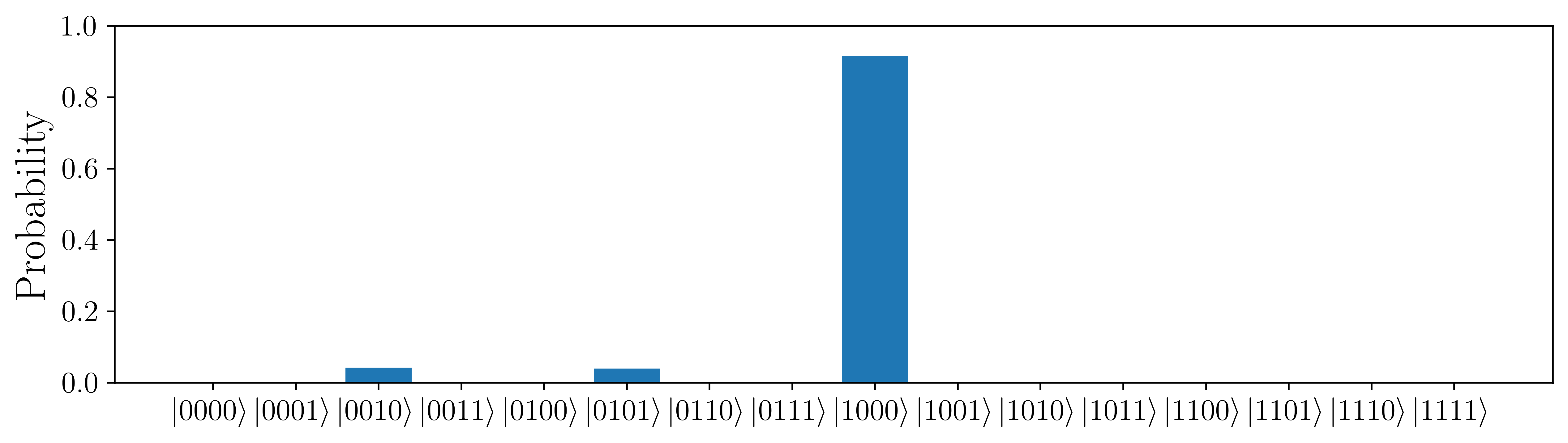}
    \caption{$H=0.1$}
    \label{}
    \end{subfigure}

    \begin{subfigure}{0.85\textwidth}
        \includegraphics[width=\textwidth]{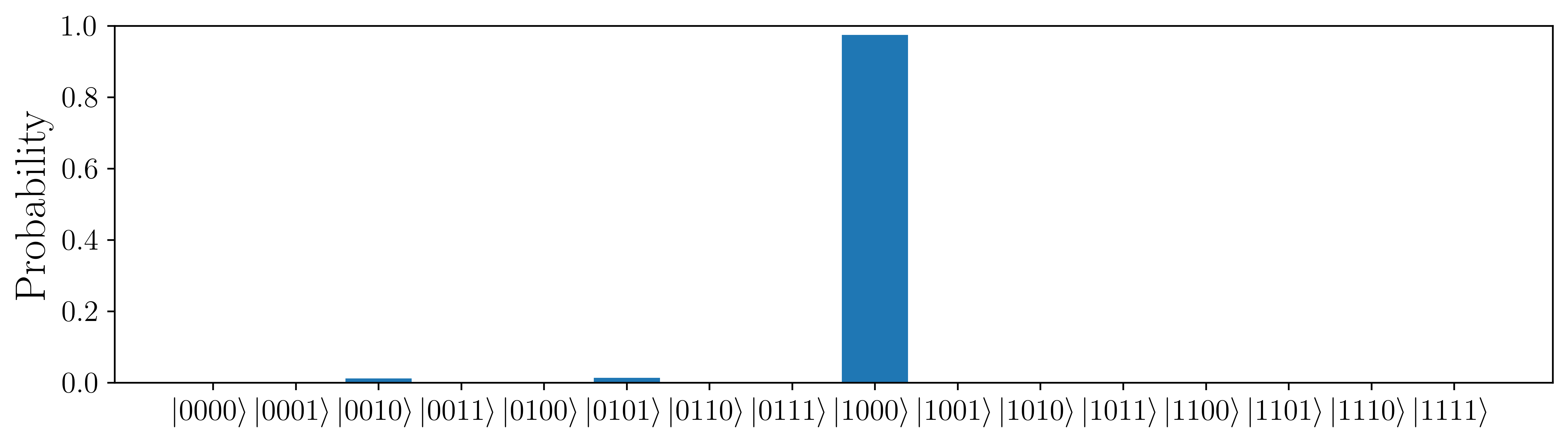}
    \caption{$H=0.01$}
    \label{}
    \end{subfigure}
    \caption{Probability distribution of $\ket{\psi(T)}$ at $T=2$ of a random 4-bit EC3 instance. Probability distributions are compared for without noise and then decreasing Hurst parameter for $H=0.3,0.2,0.1 \text{\;and\;} 0.01$.}
    \label{fig: 4bit visualizations}
\end{figure}

Next, to investigate the behavior of the probability mass function associated with the final time evolved state $\ket{\psi(T)}$, we randomly generate a 4-bit instance of EC3. We then plot the probability mass functions of the final state $\ket{\psi(T)}$ from the quantum adiabatic algorithm for varying values of the Hurst parameter at the fixed evolution time $T=2$. We take a 4-bit instance primarily due to the ease of visualization.

The instance studied in Figure \ref{fig: 4bit visualizations} is
\begin{equation}
    [(1, 2, 3), (1,3,4 ),(3,1,4),(1,2,3),(1,2,4)] \nonumber
\end{equation}
and it's unique satisfying assignment is $(1,0,0,0)$ which is easily verifiable. At $T=2$, the probability of observing $|1000\rangle$ after measuring $\ket{\psi(T)}$ is 0.257. (b),(c),(d) and (e) are probability distributions of the final state $\ket{\psi(T)}$ after single realizations of time evolution with noise driven by fractional Brownian motion with $H=0.3,0.2,0.1 \text{\;and\;} 0.01$. It is clear that the probability amplitude corresponding to the state $\ket{1000}$ is amplified as $H$ decreases and approaches 0 while the probability amplitudes corresponding to other basis states are diminished.

\subsection{An ensemble of 1000 random 8-bit instances}

Until now, we have focused our attention on some specific instances of EC3. Naturally, one may wonder if the improvements persist for all possible instances of EC3. This is of particular interest from an optimization point of view. Since it is not theoretically possible to test all the possible instances, we take a randomly generated set of 1000 8-bit instances and observe their average fidelities with varying Hurst parameter values in the range $(0,\frac{1}{2})$ at a limited evolution time of $T=2$. 

\begin{figure}
    \centering
    \includegraphics[width=0.95\linewidth]{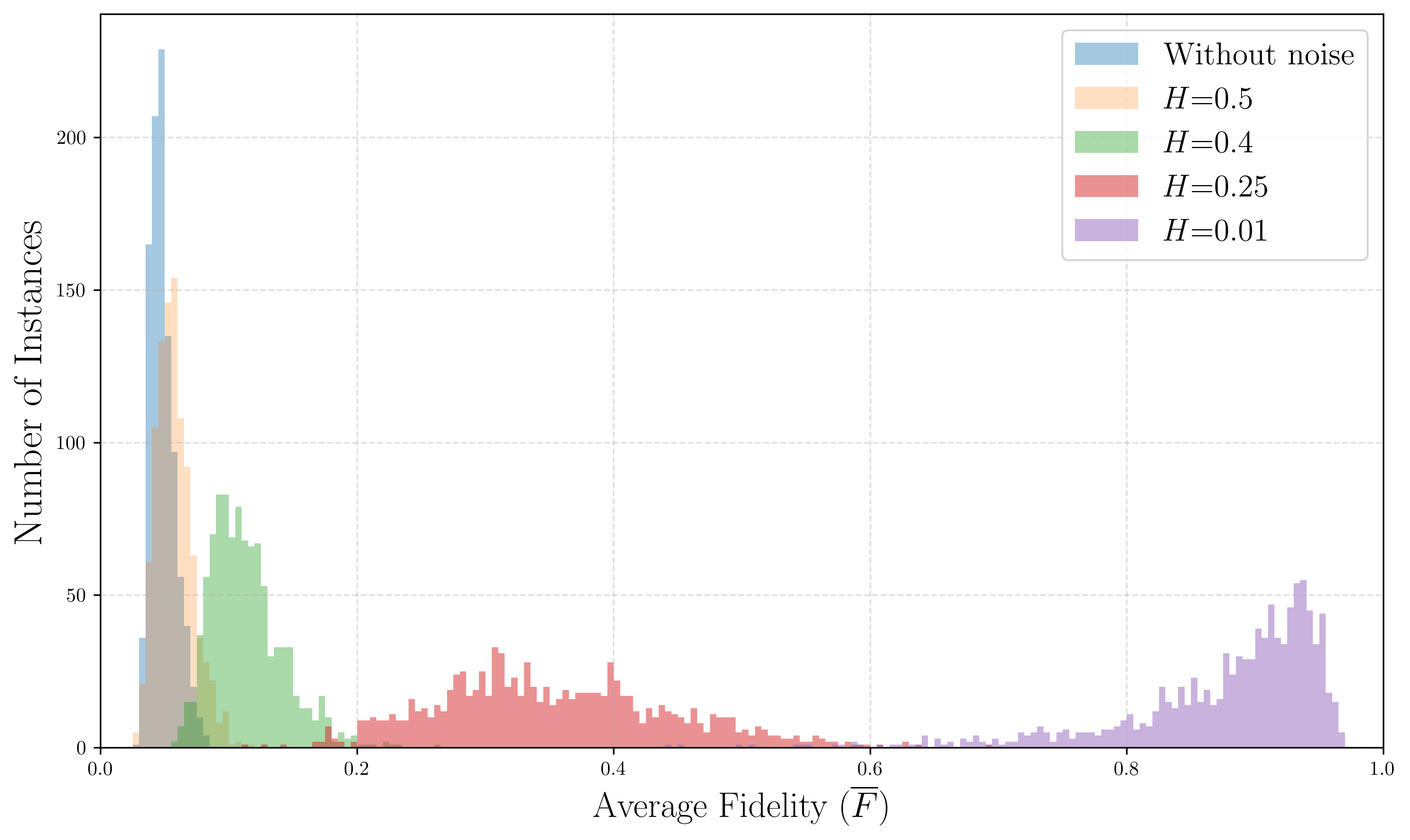}
    \caption{Histograms of average fidelities of 1000 randomly generates 8-bit EC3 instances for the standard quantum adiabatic algorithm and the quantum adiabatic algorithm in the presence of noise driven by fractional Brownian motion at $H=0.5,0.4,0.25,0.01$ drawn on top of each other.}
    \label{fig: Histogram_1000}
\end{figure}

As implied by the histogram in Figure \ref{fig: Histogram_1000}, we can answer this question mostly affirmatively. It is evident that the average fidelities generally increase for all instances considered as $H$ decreases at a fixed evolution time. However, we cannot exclude the possibility of existing an instance that does not improve its average fidelity with decreasing $H$ in the regime of $0<H<\frac{1}{2}$. It is also worth noting that to the left of Figure \ref{fig: Histogram_1000}, the histograms are concentrated and spread out as they move right with decreasing $H\in (0,\frac{1}{2})$. This emphasizes that instances with similar fidelities under the standard quantum adiabatic algorithm may exhibit significantly different but improved responses to the quantum adiabatic algorithm in the presence of noise driven by fractional Brownian motion.

\subsection{Scaling behavior}

\begin{figure}[p]
    \centering
    \begin{subfigure}
    {0.98\textwidth}
        \includegraphics[width=\linewidth]{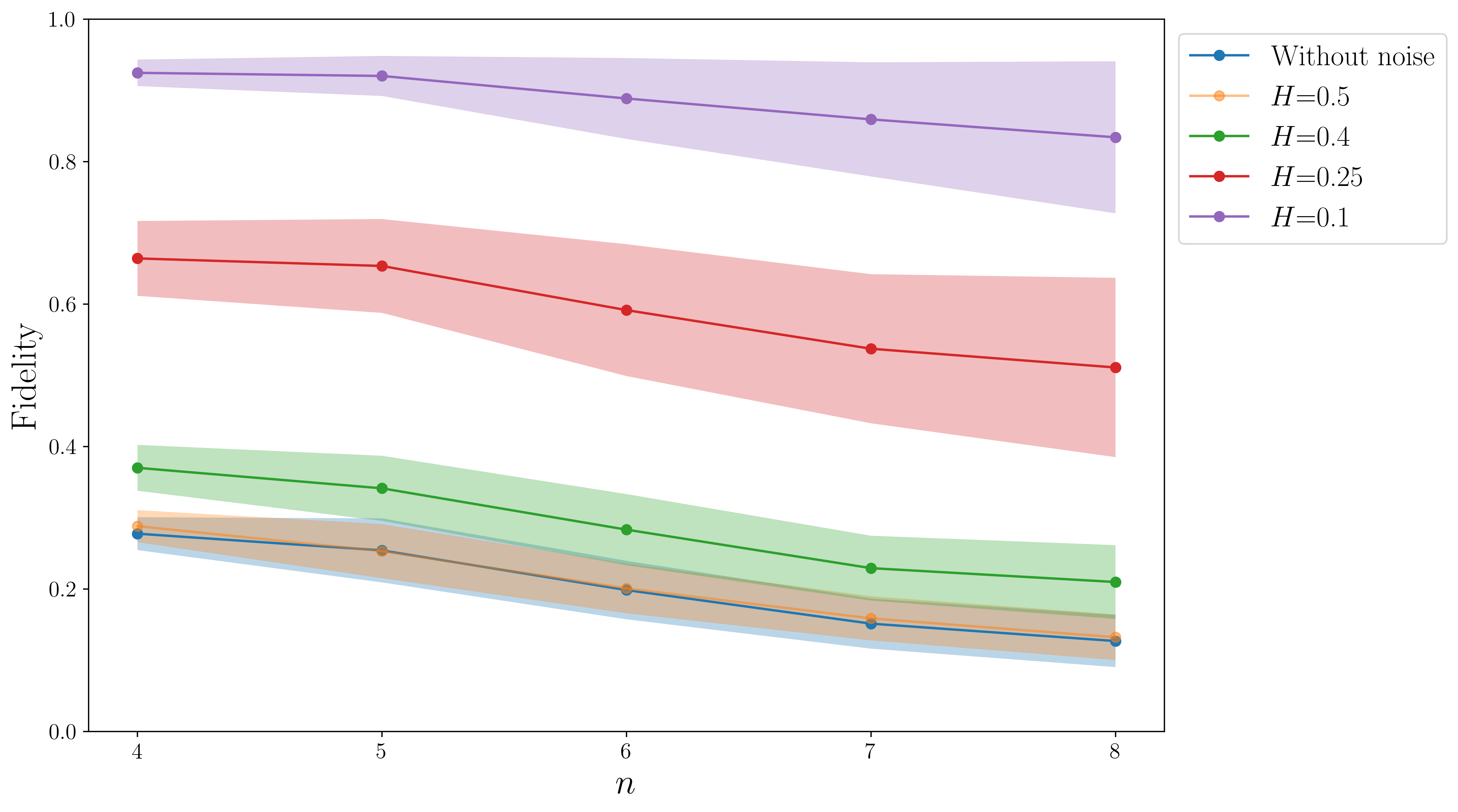}
        
        \caption{Average fidelity}
        \label{fig: Scaling plot - a}
    \end{subfigure}
    
    \vspace{1em}
    
    \begin{subfigure}
    {0.93\textwidth}
       \includegraphics[width=\linewidth]{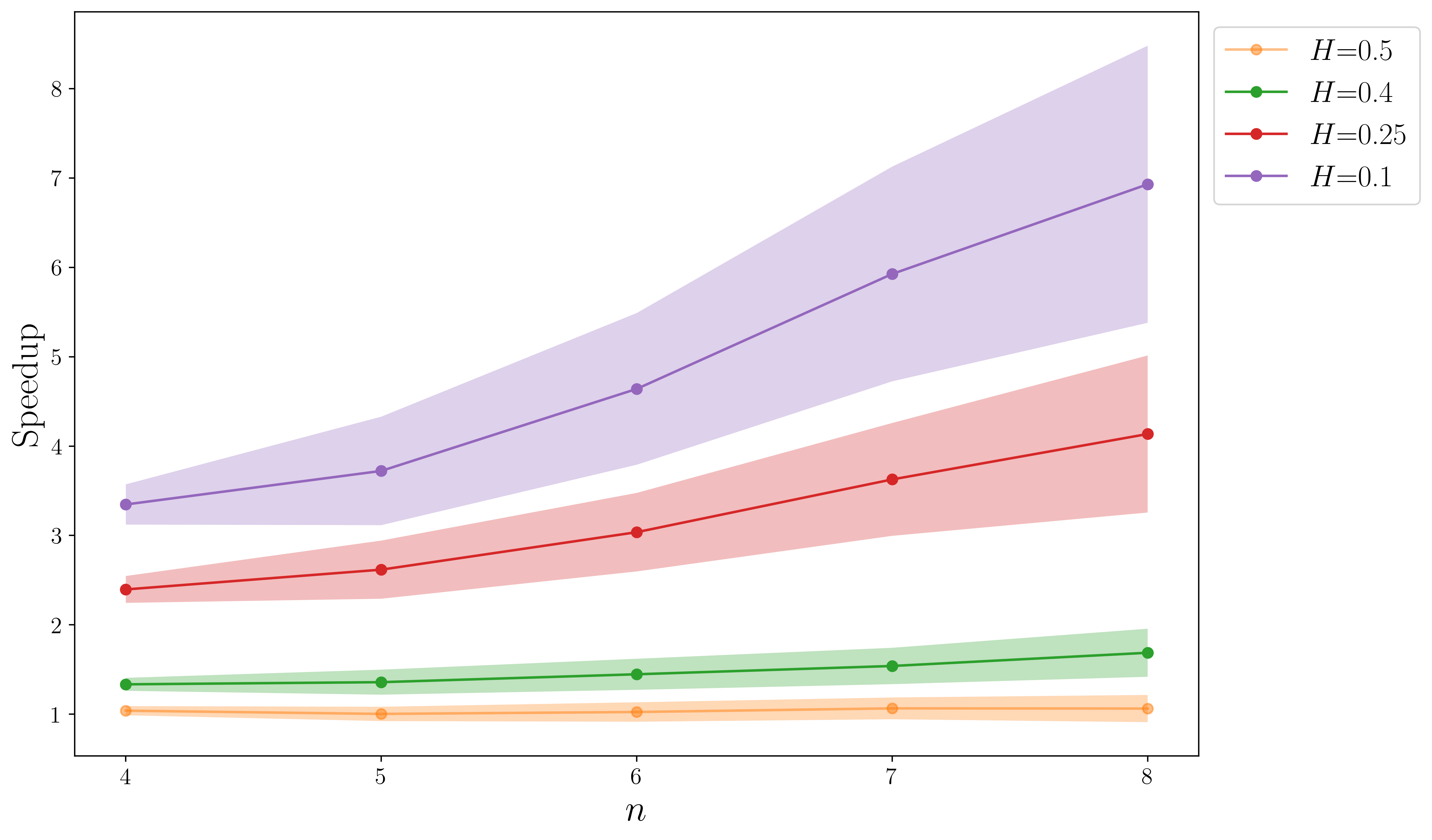}
        
        \caption{Speedup}
        \label{fig: Scaling plot - b}
    \end{subfigure}
    
    \caption{Scaling behaviors of Average fidelity and Speedup as the number of bits $n$ of an instance increases. The area covering one standard deviation from above and below are highlighted in each variation.}
    \label{fig: scaling plots}
\end{figure}

A fundamental question arising from our previous results is how these improvements scale with the system size $n$. To test this, for each $n$ from 4 to 8, we generate 100 random instances of $n$-bit EC3 instances and perform the standard quantum adiabatic algorithm and the quantum adiabatic algorithm in the presence of noise driven by fractional Brownian motion for $H=0.5,0.4,0.25 \text{\;and\;} 0.1$. We calculate the average fidelity for each instance and then plot the mean of 100 instances for each $n$. For all $n$, both standard and noisy quantum adiabatic algorithm was performed for $T=5$, and the results are presented in Figure \ref{fig: Scaling plot - a} As $n$ increases, we observe a slight decrease in average fidelities for all $H$ values. This is because as the number of bits increases, in general the fidelity achieved at a constant evolution time decreases in the quantum adiabatic algorithm. Thus, to better understand how the quantum adiabatic algorithm in the presence of noise performs with scaling, we take the speedup (SP) of the 100 instances and plot their means for each $n$. This is presented in Figure \ref{fig: Scaling plot - b}. Here we can see that for larger $H$ values, there is no visible change in speedup as $n$ increases. However, with values of $H$ 0.25 and 0.1, there is a significant increase in the speedup with increasing $n$. Even though we have simulated up to 8 bits, we can predict that the speedup will increase as the number of bits increases for higher number of bits. This is a very crucial result that indicates the scalability of the quantum adiabatic algorithm in the presence of noise driven by fractional Brownian motion.

\subsection{Evolution of a single qubit mapped on the Bloch sphere}

\begin{figure}
    \centering
    \includegraphics[width=1\linewidth]{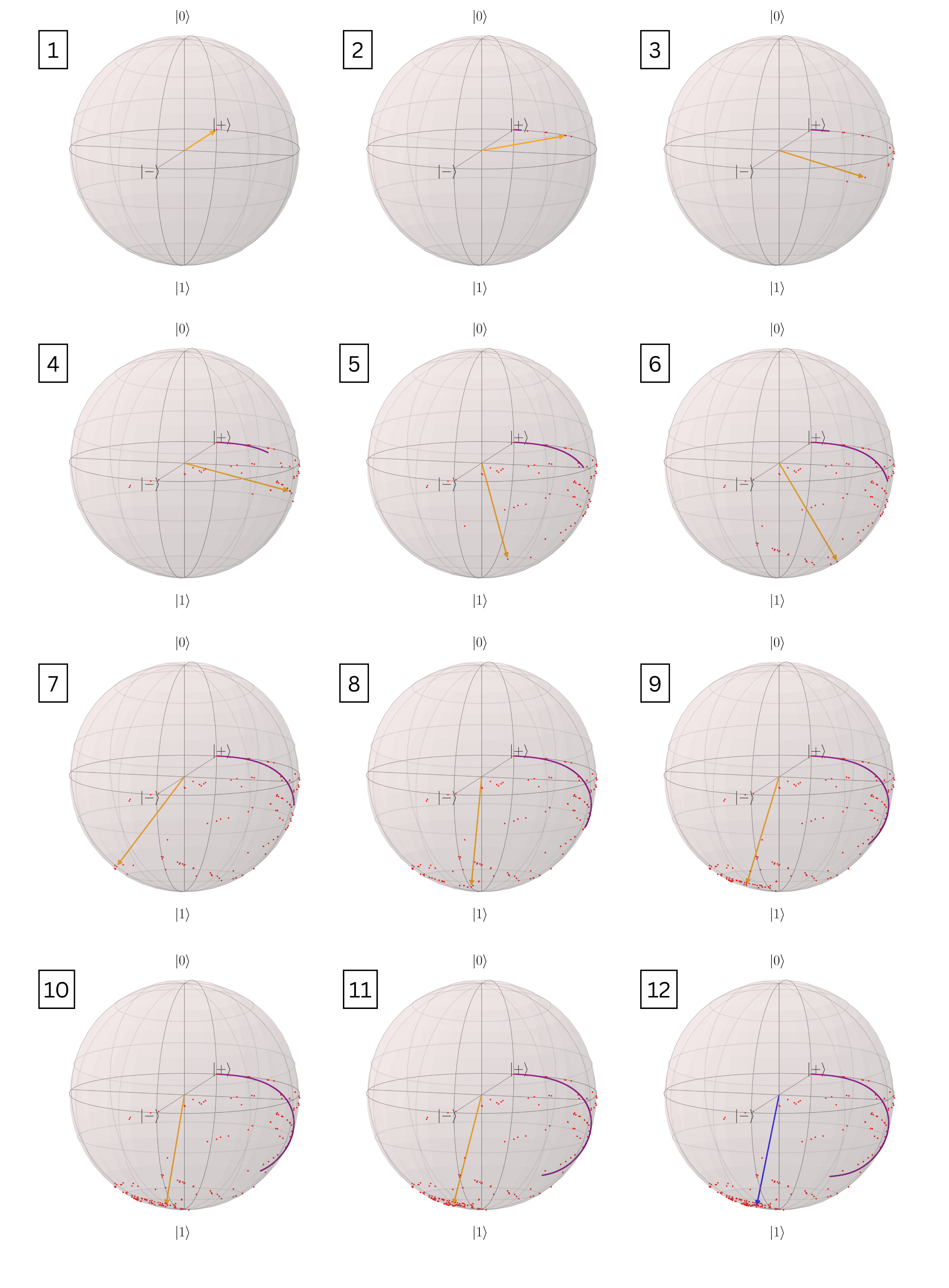}
    \caption{Snapshots of time evolution trajectories of a single qubit system with the standard quantum adiabatic algorithm (purple) and the quantum adiabatic algorithm in the presence of noise driven by fractional Brownian motion with $H=0.3$ (red). The evolution time is $T=2$. The yellow arrows point the current state of the trajectory of the noisy quantum adiabatic algorithm and the blue arrow in the final snapshot points the final state. \href{https://github.com/Osanda00/Undergraduate-thesis/blob/main/Bloch_sphere_mapping.mp4}{GitHub}}
    \label{fig: Bloch sphere plotting}
\end{figure}

Finally, to complement these findings and gain a deeper insight into the dynamics of time evolution, it would be valuable to visualize and compare the state trajectories under both the standard quantum adiabatic algorithm and the quantum adiabatic algorithm in the presence of fractional Brownian motion-driven noise. We can only do this efficiently for a two-level (one qubit) system using a Bloch sphere. For that, we consider the following final (problem) Hamiltonian
\begin{equation}
    H_F = 
    \begin{bmatrix}
        1.5 & 0 \\
        0 & 1.5
    \end{bmatrix}
\end{equation}
whose ground state is $\ket{1}$. The initial Hamiltonian in \ref{eq: initial Hamiltonian} reduces to 
\begin{equation}
    H_I = -\sigma_{x} = 
    \begin{bmatrix}
        0 & -1 \\
        -1 & 0
    \end{bmatrix}
\end{equation}
in the case of $n=1$. We start the time evolution in the ground state of $H_I$ which is $\ket{+}$ and simulations are run for an evolution time of $T=2$ for both the standard quantum adiabatic algorithm and the quantum adiabatic algorithm in the presence of noise driven by fractional Brownian motion with $H=0.3$. The time evolution paths were mapped on the Bloch sphere, and Figure \ref{fig: Bloch sphere plotting} presents snapshots of the time evolution. The final fidelity of the quantum adiabatic algorithm without noise is 0.08204 while fidelity with noise is 0.9447. Snapshots in Figure \ref{fig: Bloch sphere plotting} clearly indicates that the path taken by the quantum adiabatic algorithm with noise represented by red dots finishes more closely to the ground state of the final Hamiltonian than the quantum adiabatic algorithm without noise, represented by purple dots. Furthermore, we observe that this is achieved by taking larger steps involving randomness which is characteristic to algorithms with stochasticity. 

In summary, our numerical simulations demonstrate that the quantum adiabatic algorithm in the presence of noise driven by fractional Brownian motion with the Hurst parameter $H$ in the range $(0,\frac{1}{2})$ outperforms the standard quantum adiabatic algorithm with superior fidelity, particularly at limited evolution times. Furthermore, the speedup gained by the quantum adiabatic algorithm in the presence of fractional Brownian motion-driven noise increases with $n$. It should be noted that the quantum adiabatic algorithm in the presence of noise driven by fractional Brownian motion serves as a theoretical framework. However, its insights may be relevant to practical applications, particularly in quantum annealing, which inherently operates in noisy environments.

\section{Conclusions}
This study investigated the impact of performing adiabatic quantum computing in the presence of stochastic noise induced by fractional Brownian motion. The main objective was to examine whether incorporating noise into the adiabatic evolution could improve the success probabilities of the quantum adiabatic algorithm at limited evolution times, which has been answered positively. 

\par To test this, we first derived the stochastic schr\"{o}dinger equation with multiplicative noise driven by a semimartingale approximation of fractional Brownian motion. The quantum adiabatic algorithm in the presence of noise driven by fractional Brownian motion was benchmarked with the Exact Cover 3 problem, a well-known candidate in the NP-complete problem list. Numerical simulations were performed using a discretization of the  continuous time evolution on a classical computer up to 8 qubits.

\par The key findings of this study can be summarized as follows. Our analysis revealed that performing the quantum adiabatic algorithm in the presence of noise driven by fractional Brownian motion influenced its success probability, with the impact being highly sensitive to the Hurst parameter ($H$). In the regime of $0<H<\frac{1}{2}$, noise improved the average fidelity of the algorithm, gaining significant speedup over the standard (noiseless) quantum adiabatic algorithm at limited evolution times. These improvements become more pronounced as $H$ approaches $0$. In contrast, for $\frac{1}{2}\leq H<1$, the incorporation of noise did not have an influence on average fidelities at limited evolution times and therefore did not show a significant speedup. Perhaps most promisingly, our results demonstrate favorable scaling characteristics, as the speedup of the quantum adiabatic algorithm in the presence of noise driven by fractional Brownian motion increases with the qubit count, despite minor reductions in average fidelities.

These results suggest that if a quantum system with noise driven by fractional Brownian motion with $0<H<\frac{1}{2}$ is physically realizable, performing the quantum adiabatic algorithm in such a system would significantly improve its success probabilities.

\bibliographystyle{IEEEtran} 
\bibliography{References/references}

\end{document}